\documentstyle[aps,prl,multicol,psfig]{revtex}
\newcommand{\be}{\begin{equation}} 
\newcommand{\ee}{\end{equation}} 
\newcommand{\bea}{\begin{eqnarray}} 
\newcommand{\eea}{\end{eqnarray}} 
\newcommand{\eq}{Eq.~(\ref} 
\def\dm{\dot{\tilde{M}}_{kl}}
\def\bi{{\bf i}}              %
\def\bs{{\bf s}}              %
\def\bU{{\bf U}}              %
\def\bP{{\bf P}}              %
\def\bom{\mbox{\boldmath$ \omega $}}
\def\bv{{\bf v}}              %
\def\bx{{\bf x}}              %
\def\br{{\bf r}}              %
\def\bee{{\bf e}}              %
\def\bg{{\bf g}}              %
\def\bn{{\bf n}}              %
\def\bF{{\bf F}}              %
\def\bG{{\bf G}}              %
\def\bL{{\mathcal{L} }}              %
\def\bJ{{\bf J}}              %
\def\la{\langle}              %
\def\ra{\rangle}              %
\def\dd{\text{d}}              %
\begin{document}
\draft
\title{Foundations of Dissipative Particle 
Dynamics}
\author{Eirik G. Flekk\o y$^1$, Peter V. Coveney$^2$ 
and Gianni De Fabritiis$^2$}
\address{%ad---------------------------------------------------------
$^1$Department of Physics, University of Oslo\\
        P.O. Box 1048 Blindern, 0316 Oslo 3, Norway\\
$^2$Centre for Computational Science, Queen Mary and Westfield College,\\
        University of London, London E1 4NS, United Kingdom}
\date{\today}
\maketitle 
\begin{abstract}
%NB
We derive a mesoscopic modeling and simulation
technique that is very close to the
technique known as dissipative particle dynamics.
The model is derived  from  molecular dynamics 
by means of a systematic coarse-graining procedure. This procedure
links the forces between the dissipative particles
to a hydrodynamic description of the underlying molecular dynamics (MD)
particles.
In particular, the  dissipative particle forces are given 
directly in terms of the viscosity emergent from MD, while the 
interparticle energy transfer is similarly given by the heat 
conductivity derived from MD.
In linking the microscopic and mesoscopic descriptions we thus
rely on the macroscopic description emergent from MD.
Thus the rules governing our new
form of dissipative particle dynamics reflect the 
underlying molecular dynamics; in particular all the 
underlying conservation laws carry over from the 
microscopic to the mesoscopic descriptions.
We obtain the forces experienced 
by the dissipative particles together with an approximate 
form of the associated equilibrium distribution.
 Whereas previously the dissipative particles were spheres of 
fixed size and mass, now they
are defined as cells on a Voronoi lattice with variable masses and sizes.
This Voronoi lattice arises naturally from the coarse-graining 
procedure which may be applied iteratively
and thus represents a form of renormalisation-group mapping.
It enables us to select any desired local scale
for the mesoscopic description of a given problem. Indeed, the method 
may be used to deal with situations in which several 
different length scales are simultaneously present. We compare and 
contrast this new particulate model with existing continuum
fluid dynamics techniques, which rely on a purely macroscopic and 
phenomenological approach.
Simulations carried out with the  present scheme show 
good agreement with theoretical predictions for the equilibrium behavior.
\end{abstract}

%==================================================================
\begin{multicols}{2}

\section{Introduction}

The non-equilibrium behavior of fluids continues to present 
a major challenge for both theory and numerical simulation. In recent times, 
there has
been growing interest in the study of so-called `mesoscale' modeling and 
simulation 
methods, particularly for the description of the complex dynamical behavior of 
many kinds
of soft condensed matter, whose properties have thwarted more 
conventional approaches. 
As an example, consider the case of complex fluids with many
coexisting length and time scales, for which hydrodynamic descriptions are 
unknown and may not even exist. These kinds of fluids 
include multi-phase flows, particulate and colloidal suspensions, polymers,
and amphiphilic fluids, including emulsions and microemulsions. 
Fluctuations and
Brownian motion are often key features controlling their behavior.

{}From the standpoint of traditional fluid dynamics, a general problem
in describing 
such fluids is the lack of adequate continuum models. 
Such descriptions, which are usually based on simple
conservation laws, approach the physical description from the 
macroscopic side, that is in a `top down' manner, 
and have certainly proved successful 
for simple Newtonian fluids \cite{landau59}.
For complex fluids, however, equivalent phenomenological representations 
are usually unavailable
and instead it is necessary to base the modeling approach
on a microscopic (that is on a particulate) description of the system, 
thus working from the bottom upwards, along the general lines 
of the program for statistical mechanics pioneered by 
Boltzmann \cite{boltzmann1872}.
Molecular dynamics (MD) presents itself as the most accurate 
and fundamental method~\cite{koplik95} but it is far too
computationally intensive to provide a practical option for 
most hydrodynamic problems involving complex fluids.
Over the last decade several alternative `bottom up' strategies have
therefore been introduced. Hydrodynamic lattice gases \cite{frisch86},
which model the fluid as a discrete set of particles, represent
a computationally efficient spatial and temporal discretization of the 
more conventional molecular dynamics. The
lattice-Boltzmann method \cite{mcnamara88}, originally derived from 
the lattice-gas paradigm by invoking Boltzmann's {\em Stosszahlansatz}, 
represents an 
intermediate (fluctuationless) approach 
between the top-down (continuum) and bottom-up (particulate)
strategies, insofar as the basic
entity in such models is a single particle distribution function; but
for interacting systems even these lattice-Boltzmann methods can be 
subdivided into bottom-up~\cite{chan93} 
and top-down models~\cite{swift95}.

A recent contribution to the family of bottom-up approaches
is the dissipative particle dynamics (DPD) method introduced
by Hoogerbrugge and Koelman in 1992~\cite{hoogerbrugge92}.
Although in the original formulation
of DPD time was discrete and space continuous, a more recent re-interpretation 
asserts that
this model is in fact a finite-difference approximation to the `true'
DPD, which is defined by a set of continuous time 
Langevin equations with momentum
conservation between the dissipative particles~\cite{espanol95b}.
Successful applications of the technique have been made to colloidal
suspensions~\cite{boek97}, polymer solutions~\cite{schlijper95} and
binary immiscible fluids~\cite{coveney96}.
For specific applications where comparison is possible,
this algorithm is orders of magnitude faster than MD~\cite{groot97}.
The basic elements of the DPD scheme are particles that 
represent rather ill-defined `mesoscopic' quantities of the 
underlying molecular fluid. 
These dissipative particles are stipulated to evolve in the 
same way that MD 
particles do, but with different inter-particle forces: since 
the DPD particles are pictured to have 
internal degrees of freedom, the forces between them have 
both a fluctuating and a dissipative component in addition
to the conservative forces that are present at the MD level.
Newton's third law is still satisfied, however, and consequently 
momentum conservation together with mass conservation produce 
hydrodynamic behavior at the macroscopic level.

Dissipative particle dynamics has been  shown to produce the
correct macroscopic (continuum) theory; that is, for a one-component DPD
fluid, the Navier-Stokes equations emerge in the 
large scale limit, and the fluid viscosity 
can be computed~\cite{espanol95,MBE1}.
%NB
However, even though dissipative particles have
generally   been viewed as  clusters of molecules, 
no attempt has been made to link DPD to the underlying 
microscopic dynamics, and DPD thus remains a foundationless
algorithm, as is that of the hydrodynamic lattice gas and {\em a
fortiori} the lattice-Boltzmann
method. It is the principal purpose of the present paper
to provide an atomistic foundation for dissipative
particle dynamics. Among the numerous benefits gained by achieving
this, 
we are then able to provide a precise definition of the term `mesoscale', 
to relate the hitherto purely
phenomenological parameters in the algorithm to underlying molecular 
interactions, and thereby to formulate DPD simulations for {\em specific} 
physicochemical systems, defined in terms of their molecular constituents.
The DPD that we derive is a representation of the underlying MD.
Consequently, to the extent that the 
approximations made are valid,  the DPD and MD will have the same hydrodynamic 
descriptions,
and no separate kinetic theory for, say, the DPD viscosity
will be needed once it is known for the MD system.
Since the MD degrees of freedom will be integrated out in our
approach the MD viscosity will appear in the DPD model as 
a parameter that may be tuned freely.

In our approach, the `dissipative particles' (DP) are defined in terms
of appropriate weight functions that sample portions of the underlying 
conservative MD particles, and the 
forces between the dissipative particles are obtained from the 
hydrodynamic 
description of the MD system: the microscopic conservation laws carry 
over directly to the 
DPD, and the hydrodynamic behavior of MD is thus reproduced by the 
DPD, albeit at a coarser scale. The mesoscopic (coarse-grained) scale
of the DPD can be precisely
specified in terms of the MD interactions.
The size of the dissipative particles, as specified by 
the number of MD particles within them, furnishes the 
meaning of the term `mesoscopic' in the present context.
Since this size is a freely tunable parameter of the model, the
resulting DPD
introduces a general procedure for simulating 
microscopic systems at 
any convenient scale of coarse graining, provided that
the forces between the dissipative particles are known.
When a hydrodynamic description of the underlying particles 
can be found, these forces follow directly; in
cases where this is not possible, the forces between dissipative
particles must be supplemented 
with the additional components of the physical description 
that enter on the mesoscopic level.

The DPD model which we derive from molecular dynamics
is formally similar to conventional, albeit foundationless, 
DPD~\cite{espanol95}. 
The interactions are pairwise and conserve mass and momentum, as
well as energy \cite{avalos97,espanol97}.
Just as the forces conventionally used to define DPD have
conservative, dissipative and fluctuating 
components, so too do the forces in the present case. 
In the present model, the role of the conservative force is played 
by the pressure forces.
However, while conventional dissipative particles possess spherical
symmetry and experience interactions mediated by purely central
forces, our dissipative particles 
are defined as space-filling cells on a Voronoi lattice whose forces
have both central and tangential components.
These features are shared with a model studied by Espa\~{n}ol
\cite{espanol98b}. This model links DPD to smoothed particle 
hydrodynamics \cite{monaghan92} and defines the DPD 
forces by hydrodynamic considerations in a way analogous to earlier
DPD models.
Espa\~{n}ol {\it et al.} \cite{espanol97b} have also carried out MD
simulations with a superposed Voronoi mesh 
in order to measure the coarse grained inter-DP forces.

While conventional DPD defines dissipative particle masses
to be constant, this feature is not preserved in our new model. 
In our first publication on this theory \cite{flekkoy99}, we 
stated that, while the dissipative particle masses fluctuate due to the
motion of MD particles across their boundaries, the average masses
should be constant. In fact, the DP-masses
vary due to distortions of the Voronoi cells, 
and this feature is now properly incorporated
in the model.

We follow two distinct routes to obtain the
fluctuation-dissipation relations that 
give the  magnitude of the thermal forces.
The first route follows the conventional path which makes
use of a Fokker-Planck equation~\cite{espanol95b}.
 We show that  the DPD system is described in an approximate 
sense by the isothermal-isobaric ensemble.
The second route is based on the theory of
fluctuating hydrodynamics and it is argued that this
approach corresponds to the statistical mechanics of
the grand canonical ensemble. 
Both routes lead to the same result for the fluctuating 
forces and simulations confirm that, 
with the use of these forces, the measured 
DP temperature is equal to the MD temperature which is provided as input.
This is an important finding in the present context as
the most significant 
approximations we have made 
underlie the derivation of the thermal forces.

\section{Coarse-graining molecular dynamics: from micro to mesoscale}

The essential idea motivating our definition of mesoscopic 
dissipative particles is to specify them as clusters of MD particles
in such a way that the MD particles themselves remain unaffected while
{\em all} being represented by the dissipative particles. 
The independence of the molecular dynamics from the superimposed
coarse-grained dissipative particle dynamics implies that 
the MD particles are able to move between the dissipative particles.
The stipulation that all MD particles must be fully represented by the
DP's implies that while the mass, momentum and energy
of a single MD particle may be shared between DP's, the sum of the
shared components must always equal the mass and momentum of the MD particle.

\subsection{Definitions}
Full representation of all the MD particles can be achieved in a
general way by introducing a sampling function
\be
f_k(\bx )= \frac{s(\bx - \br_k )}{\sum_l s(\bx - \br_l )}
\label{sampling} \; .
\ee
where the positions $\br_k$ and $\br_l$ define the DP 
centers, $\bx$ is an arbitrary position and $s(\bx )$
is some localized function. 
It will prove convenient to choose it as a Gaussian 
\be 
s(\bx ) = \exp{( - x^2/a^2)}
\ee
where the distance $a$ sets the scale of the sampling function,
although this choice is not necessary.
The mass, momentum and internal energy $E$ of the $k$th DP are then 
defined as
\bea
M_k &=& \sum_i   f_k(\bx_i ) m , \nonumber \\
\bP_k &=& \sum_i  f_k(\bx_i )m \bv_i , \nonumber \\
\frac{1}{2} M_k U_k^2 + E_k &=& \sum_i f_k(\bx_i )
 \left( \frac{1}{2} m v_i^2 + \frac{1}{2} \sum_{j\neq i} 
V_{MD}(r_{ij}) \right) \nonumber \\
& \equiv & \sum_i f_k(\bx_i ) \epsilon_i ,   \label{DPdef}
\eea
where $\bx_i$ and $\bv_i$ are the position and velocity 
of the $i$th MD particle, which are all assumed to have identical masses $m$, 
$\bP_k$ is the momentum 
of the $k$th DP and $V_{MD}(r_{ij})$ is the  potential energy 
of the MD particle pair $ij$, separated a distance $r_{ij}$.
The particle energy $\epsilon_i$ thus contains both the kinetic
and a potential term.
The kinematic condition
\be
\dot{\br}_k =\bU_k \equiv \bP_k/M_k 
\ee
completes the definition of our dissipative particle dynamics.

It is generally true that mass and momentum conservation
suffice to produce hydrodynamic behavior. However, 
the equations expressing these conservation laws contain the
fluid pressure. In order to get the fluid pressure a 
thermodynamic description of the system is needed.
This produces an equation of state, which
closes the system of hydrodynamic equations.
Any thermodynamic potential may be used to obtain
the equation of state.
In the present case we shall take this potential to be 
the internal energy $E_k$ of the dissipative particles,
and we shall obtain the equations of motion
for the DP mass, momentum and energy.
Note that the internal energy would also have to be computed
if a free energy had been chosen for the thermodynamic
description. For this reason it is not possible to complete
the hydrodynamic description without taking the
energy flow into account. As a byproduct of this 
the present DPD also contains a description
of the heat flow and corresponds to the recently
introduced DPD with energy conservation~\cite{avalos97,espanol97}.
Espa\~{n}ol previously introduced an angular momentum variable
describing the dynamics of extended particles~\cite{espanol98b}: this
is needed
when forces are non-central 
in order to avoid dissipation of energy in a rigid rotation of the fluid.
Angular momentum could be included on the same footing as momentum 
in the following developments.
However for reasons both of space and 
conceptual economy we shall omit it in the present context, even
though it is probably important in applications
where hydrodynamic precision is important.
% be seen by considering a vortex enclosed in angular momentum 
% conserving boundaries. Within such boundaries both a system 
% of DP's with and without an internal angular momentum variable will
% produce an angular momentum conserving flow, the difference
% being that the flow without the angular momentum variables 
% will not relax to a rigid body rotation but rather a vortex 
% with some internal shear (though without dissipation).
In the following sections, we shall 
use the notation $\br$, $M$, $\bP$  and $E$ with the 
indices $k\; , l\; , m$ and $ n$ to denote DP's
while we shall use $\bx$, $m$, $\bv$ and $\epsilon$ 
with the indices $i$ and $j$ to denote MD particles.

\subsection{Equations of motion for the dissipative particles
based on a microscopic description} 
The fact that all the MD particles are represented at all instants in the
coarse-grained scheme is guaranteed by the normalization condition 
$\sum_k f_k (\bx ) = 1$. This implies directly that 
\bea
\sum_k M_k  &=& \sum_i m \nonumber \\
\sum_k  \bP_k &=& \sum_i m \bv_i \nonumber \\
\sum_k E_k^{\text{tot}}  &=&
\sum_k \left( \frac{1}{2} M_k \bU_k^2 + E_k \right) = \sum_i \epsilon_i \;;
\eea
thus with mass, momentum and energy conserved at the MD 
level, these quantities are also conserved at the DP level.
In order to derive the equations of motion for
dissipative particle dynamics we now take the time derivatives 
of Eqs.~(\ref{DPdef}).
This gives
\bea
\frac{\dd {M_k}}{\dd t} &=& \sum_i  \dot{f}_k (\bx_i )m \label{mass}\\
\frac{\dd  \bP_k }{\dd t} &=& \sum_i \left( \dot{f}_k  (\bx_i ) m \bv_i
+{f}_k (\bx_i ) \bF_i \right)  \label{momentum} \\
\frac{\dd E_k^{\text{tot}}}{\dd t}   &=&\sum_i \left( \dot{f}_k  (\bx_i ) \epsilon_i
+{f}_k (\bx_i ) \dot{\epsilon}_i \right)  \label{energy} 
\eea
where $\dd /\dd t$ is the substantial derivative and $\bF_i = m\dot{\bv}_i$
is the force on particle $i$.

The Gaussian form of $s$ implies that \\ $\dot{s} (\bx ) = -(2/a^2)
\dot{\bx} \cdot \bx s(\bx )$. This makes it possible to write
\be
\dot{f}_k (\bx_i ) = f_{kl}(\bx_i ) (\bv'_i \cdot \br_{kl}
+ \bx'_i \cdot \bU_{kl} )
\label{dotf}
\ee
where the overlap function $f_{kl}$ is defined as
$f_{kl} (\bx )\equiv (2/a^2 )f_{k} (\bx ) f_{l} (\bx ) $,
$\br_{kl} \equiv (\br_k - \br_l)$ and $\bU_{kl} \equiv (\bU_k - \bU_l)$,
and we have rearranged terms so as to get them in terms of the 
centered variables
\bea 
\bv'_i &=& \bv_i - \frac{(\bU_k +\bU_l)}{2} \nonumber \\
\bx'_i &=& \bx_i - \frac{(\br_k +\br_l)}{2}  \; .
\label{primes}
\eea
%****** VORONOI STORY*******YY

Before we proceed with the derivation of the 
equations of motion it is instructive 
to work out the actual forms of $f_k(\bx )$
and $f_{kl}(\bx )$
in the case of only two particles $k$ and $l$.
 Using the Gaussian choice of $s$ we immediately get 
\be
f_k (\bx ) = \frac{1}{1+ \left[ \exp{((\bx - (\br_k + \br_l)/2)
\cdot (\br_{kl})/(a^2) )}\right]^2} \; .
\ee 
The overlap function similarly follows:
\be 
f_{kl} (\bx ) =
\frac{1}{2a^2} \cosh^{-2}{ \left( \left( \bx - \frac{\br_k + \br_l}{2} \right) 
\cdot \left( \frac{\br_{kl}}{a^2} \right) \right) }
\label{overlap} \; .
\ee
\begin{figure} 
\centerline{\hbox{\psfig{figure=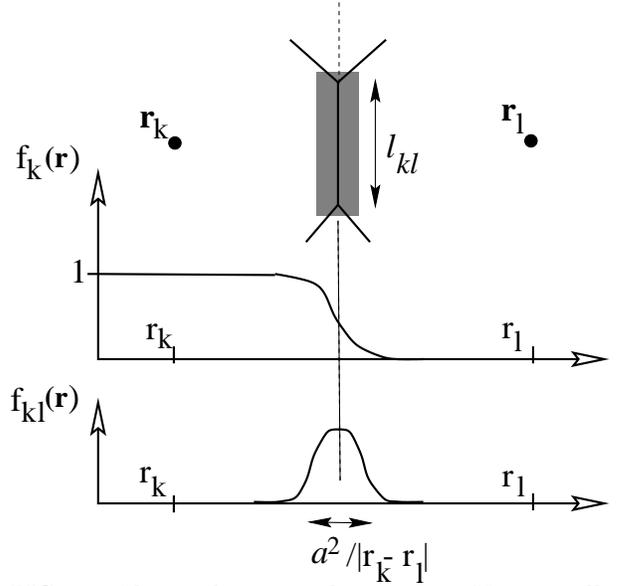,width=8cm}}}
\caption{\label{fig2}
\protect \narrowtext 
The overlap region between two Voronoi 
cells is shown in grey. The sampling function $f_k (\br )$ is shown in the
top graph and the overlap function $f_{kl} ( \br ) = 
(2/a^2) f_{k} (\br ) f_{l}(\br )$ in the bottom graph. The width of the overlap
region is $ a^2 / |\br_k - \br_l |$ and its length is denoted by $l$.}
\end{figure}
These two functions are shown in Fig.\ref{fig2}.
Note that the scale of the overlap region is not $a$  but $a^2/|\br_k - \br_l|$.
Dissipative particle interactions only
take place where the overlap function is non-zero.
This happens along the dividing line which is equally far 
from the two particles.
The contours of non-zero $f_{kl}$
thus define a Voronoi lattice with lattice segments of length 
$l_{kl}$.
This Voronoi construction is shown in Fig.~\ref{fig1}
in which MD particles in the overlap region defined by $f_{kl} >0.1$,
are shown, though presently not actually simulated as dynamic
entities.  The volume of the Voronoi cells will in general 
vary under the dynamics. 
However, even with arbitrary dissipative particle
motion the cell volumes will approach zero only exceptionally, 
and even then the identities
of the DP particles will be preserved so that they subsequently re-emerge.

\subsubsection{Mass equation}

The mass  equation (\ref{mass}) 
takes the form 
\be
\frac{\dd {M_k}}{\dd t} \equiv \sum_l \dot{M}_{kl}
\nonumber 
\ee
where
\be  \dot{M}_{kl} = \sum_i  {f}_{kl} (\bx_i )m (
   \bv_i'\cdot \br_{kl} + \bx'_i \cdot \bU_{kl}   )
\label{mass2} \; .
\ee
The $\bv'_i$ term will be shown to be negligible
within our approximations. The $ \bx'_i \cdot \bU_{kl}$-term 
however describes the geometric
effect that the Voronoi cells do not conserve their 
volume: The relative motion of the DP centers causes the cell boundaries
to change their orientation. We will return to give this `boundary 
twisting' term a quantitative content when the equations of motion
are averaged--an effect which was overlooked in our 
first publication of this 
theory \cite{flekkoy99} where it was stated that $\la \dot{M}_{kl} \ra = 0$.

\begin{figure}
\centerline{\hbox{\psfig{figure=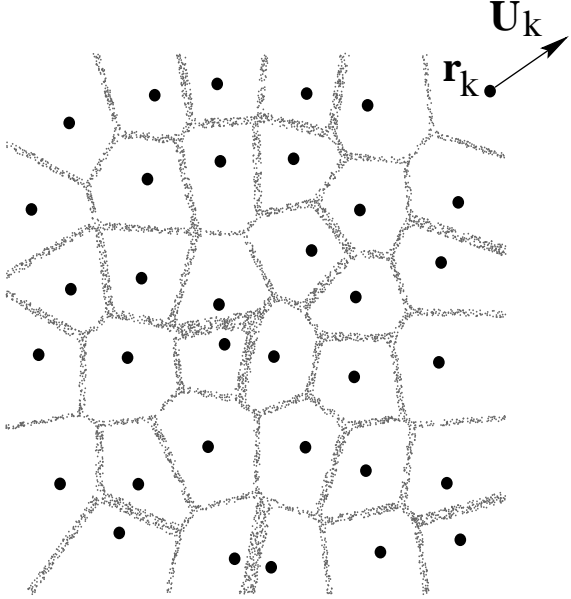,width=8cm}}}
\caption{\label{fig1}
\protect \narrowtext The Voronoi lattice defined by the 
dissipative particle positions $\br_k$. The grey dots 
which represent the underlying MD particles are drawn 
only in the overlap region.}
\end{figure}

\subsubsection{Momentum  equation}

The  momentum equation (\ref{momentum}) takes the form 
\bea
\frac{\dd  \bP_k }{\dd t} &=& 
   \sum_{li}  {f}_{kl} (\bx_i )m \bv_i (
   \bv_i'\cdot \br_{kl} + \bx'_i \cdot \bU_{kl}   )
 \nonumber \\ 
& +&  \sum_{li} f_k (\bx_i ) \bF_i 
   \label{momentum2}
\eea
We can write the force as $\bF_i = m{\bf g} +  \sum_j \bF_{ij}$
where the first term is an external force and the second term
is the internal force caused by all the other particles.
Newton's third law then takes the form $\bF_{ij} = -  \bF_{ji}$.
The last term in \eq{momentum2}) may then be rewritten as
\be
\sum_i  f_k (\bx_i )  \bF_i  =
M_k \bg + \sum_{ij}  f_k (\bx_i ) \bF_{ij}
\label{gravity}
\ee
where 
\bea
\sum_{ij}  f_k (\bx_i ) \bF_{ij} &=& - \sum_{ij}  f_k (\bx_i ) \bF_{ji} 
\nonumber \\
&=& - \sum_{ij}  f_k (\bx_j + \Delta \bx_{ij} ) \bF_{ji} \nonumber \\
&\approx &    - \sum_{ij} f_k (\bx_j ) \bF_{ji} - 
         \sum_{ij}  \left(\Delta \bx_{ij} \cdot \nabla 
f_k (\bx_i ) \right) \bF_{ji} \nonumber \\
&=&         -\frac{1}{2} \sum_{ij}  \left(\Delta \bx_{ij} \cdot \nabla 
f_k (\bx_i ) \right) \bF_{ji} \nonumber \\
&=& \sum_l \left\{ \sum_{ij} \frac{1}{2}  
f_{kl} (\bx_i ) \bF_{ij} \Delta \bx_{ij}  \right\} \cdot \br_{kl}
\label{virial}
\eea
where $\Delta \bx_{ij} = \bx_i - \bx_j$,
we have Taylor expanded $f_k (\bx )$ around $\bx_j $ and used a result
similar to \eq{dotf}) to evaluate $\nabla f_k (\bx )$.
In passing from the third to the fourth line in the above
equations we have moved the first term on the right hand side
to the left hand side and divided by two.
Now, if we group the last term above with the $\br_{kl}$ term in 
\eq{momentum2}), make use of \eq{primes}), and do some rearranging 
of terms we get 
\bea
\frac{\dd  \bP_k }{\dd t} &=&  M_k \bg +
\sum_l  \dot{M}_{kl} \frac{ \bU_{k}+\bU_{l}}{2}  \nonumber \\
&+&       \sum_{li}  f_{kl}(\bx_i ){\bf \Pi}_i'  
 \cdot \br_{kl} \nonumber \\
&+&       \sum_{li}  f_{kl} (\bx_i ) m \bv_i' \bx_i'  \cdot \bU_{kl} 
\label{momentum3}
\eea
where we have used the relation  
$\dot{M}_k = \sum_l \dot{M}_{kl}$ and defined
the general momentum-flux tensor
\be
{\bf \Pi}_i = m \bv_i \bv_i + \frac{1}{2} \sum_{j}
\bF_{ij} \Delta \bx_{ij} \label{momentum_flux} \;.
\ee
This tensor is the momentum analogue of the mass-flux vector
$m\bv_i$. The prime indicates that the velocities on the right hand 
side are those defined in \eq{primes}).
The tensor ${\bf \Pi}_i$ describes both the momentum that the particle
carries around through its own motion and the momentum exchanged
by inter-particle forces. It may be arrived at by 
considering the momentum transport 
across imaginary cross sections of the volume in which the particle is 
located. 

%*****ENERGY EQUATION: ****************

\subsubsection{Energy equation}
In order to get the microscopic energy equation of motion 
we proceed as with the mass and momentum equations
and the two terms that appear on the right hand side 
of  \eq{energy}).

Taking $V_{MD}$ to be a central potential and using the relations 
$\nabla V_{MD}(r_{ij}) = V_{MD}'(r_{ij}) \bee_{ij} = -\bF_{ij}$ and
$\dot{V}_{MD}(r_{ij}) = V_{MD}'(r_{ij}) \bee_{ij}\cdot \bv_{ij}   
= -\bF_{ij}\cdot \bv_{ij}$ where $\bv_{ij} = \bv_i - \bv_j$ we get the 
time rate of change of the particle energy
\be
\dot{\epsilon_i } = m\bg \cdot \bv_i + \frac{1}{2} \sum_{j \neq i}
\bF_{ij} \cdot (\bv_i + \bv_j ) \; .
\ee
This gives the 
first term of \eq{energy}) in the form 
\be
\sum_i f_k (\bx_i ) \dot{\epsilon} =  \bP_k \cdot \bg
+ \frac{1}{2}  \sum_{i\neq j} f_k (\bx_i ) \bF_{ij} \cdot (\bv_i + \bv_j ) \; .
\ee
The last term of this equation is odd under the exchange $i \leftrightarrow j$
and exactly the same manipulations as in 
\eq{virial}) may be used to give
\bea
\sum_i f_k (\bx_i ) \dot{\epsilon} &=& \bP_k \cdot \bg \nonumber \\ 
& +&  \sum_{l,i\neq j} f_{kl} (\bx_i ) \frac{1}{4}  
\bF_{ij} \cdot (\bv_i + \bv_j )  \Delta \bx_{ij} \cdot \br_{kl} \nonumber \\ 
&=& \bP_k \cdot \bg
+ \sum_{l,i\neq j} f_{kl} (\bx_i ) 
\left( \frac{1}{4}  
\bF_{ij} \cdot (\bv'_i + \bv'_j )
   \right. \nonumber \\
 &+& \left. \frac{1}{2}  
\bF_{ij} \cdot \frac{\bU_k + \bU_l}{2}  \right) 
\Delta \bx_{ij} \cdot \br_{kl} 
\label{virial2}
\eea
where for later purposes  we have used  Eqs.~(\ref{primes}) 
to get the last equation.
The last term of \eq{energy}) is easily written down using
\eq{dotf}). This gives
\be
\sum_i \dot{f}_k(\bx_i) \epsilon_i  = 
\sum_{li} f_{kl} (\bx_i ) (\bv'_i\cdot \br_{kl} + \bx'_i \cdot 
\bU_{kl} )\epsilon_i \; . \label{energyA}
\ee
As previously we write the particle velocities in terms 
of $\bv_i'$. The corresponding expression for the 
particle energy is $\epsilon_i = \epsilon'_i + m\bv_i' \cdot
(\bU_k + \bU_l)/2 + (1/2) m ((\bU_k + \bU_l)/2 )^2 $ where 
the prime in $\epsilon'_i$ denotes that the particle velocity
is $\bv'_i$ rather than $\bv_i$.
Equation (\ref{energyA}) may then be written 
\bea
\sum_i \dot{f}_k(\bx_i) \epsilon_i  &= &
 \sum_l \frac{1}{2}\dot{M}_{kl} \left( \frac{\bU_k +\bU_l}{2} \right)^2 
\nonumber \\ 
&+& \sum_{li} f_{kl} (\bx_i )
\left( \epsilon'_i \bv_i' + m \bv_i' \bv_i' \cdot \frac{\bU_k +\bU_l}{2}\right)
\cdot \br_{kl} \nonumber \\
&+&  \sum_{li} f_{kl} (\bx_i ) \epsilon_i \bx'_i \cdot \bU_{kl} \; .
\eea
Combining this equation with \eq{virial2}) we obtain
\bea
\dot{E}_k^{\text{tot}} &=& \sum_{li}
 f_{kl} (\bx_i)\left(  \bJ_{\epsilon i}'
+ \Pi'_i \cdot \frac{\bU_k +\bU_l}{2} \right)
\cdot \br_{kl} \nonumber \\
&+&  M_k \bU_k \cdot \bg +
\sum_l \frac{1}{2} \dot{M}_{kl}\left( \frac{\bU_k +\bU_l}{2} \right)^2
\nonumber \\
&+& \sum_{li}    f_{kl} (\bx_i) \left( \epsilon'_i 
+ m \bv'_i \cdot \left( \frac{\bU_k + \bU_l}{2} \right)
 \right) \bx'_i \cdot \bU_{kl} \; .
\label{aa}
\eea
where the momentum-flux tensor is
defined in \eq{momentum_flux}) and  
we have identified the energy-flux vector associated 
with a particle $i$ 
\be
\bJ_{\epsilon i} = \epsilon_i \bv_i + \frac{1}{4} \sum_{i\neq j}
\bF_{ij} \cdot (\bv_i + \bv_j) \Delta \bx_{ij} \; . 
\label{enrgy_flux}
\ee
Again the prime denotes that the velocities are 
$\bv'_i$ rather than $\bv_i$.
To get the internal energy $\dot{E}_k$ 
instead of $\dot{E}_k^{\text{tot}}$
we note that $\dd   (\bP^2_k/2M_k)/\dd t  = \bU_k \cdot \dot{\bP}_k 
- (1/2) \dot{M}_k \bU^2_k$. 
Using this relation, the momentum equation
\eq{momentum3}), as well as the 
substitution $(\bU_k + \bU_l)/2 = \bU_k - \bU_{kl}/2$
in \eq{aa}), followed by some rearrangement of the $\dot{M}_{kl}$
terms we find that
\bea
\dot{E}_k^{\text{tot}} &=&
 \frac{\dd }{\dd t} \left( \frac{1}{2} M_k \bU_k^2 \right) \nonumber \\
&+& \sum_l \frac{1}{2} \dot{M}_{kl} \left( \frac{\bU_{kl}}{2} \right)^2
+ \sum_{li} f_{kl} (\bx_i)\left(  \bJ_{\epsilon i}'
- \Pi'_i \cdot \frac{\bU_{kl}}{2} \right)
\cdot \br_{kl} \nonumber \\
&+& \sum_{li}   f_{kl} (\bx_i)\left(  \epsilon'_i - m\bv'_i \cdot 
\frac{\bU_{kl}}{2} \right) \bx'_i \cdot \bU_{kl}  \; .
\label{energy_micro}\eea

This equation has a natural physical interpretation. The first 
term represents the translational kinetic energy of the DP as 
a whole. The remaining terms represent the internal energy $E_k$.
This is a purely thermodynamic quantity which cannot depend
on the overall velocity of the DP, i.e. it must be Galilean
invariant. This is easily checked as the relevant terms
all depend on velocity differences only.

The $\dot{M}_{kl}$ term represents
the kinetic energy received through mass exchange with
neighboring DPs. As will become evident when 
we turn to the averaged description, 
the term involving the momentum and energy 
fluxes represents the work done on the DP by its neighbors
and the heat conducted from them.
The $\epsilon'_i$-term represents the energy
received by the DP due to the same `boundary twisting'
effect that was found in the mass equation.
Upon averaging, the last term proportional 
to $\bv_i'$  will be shown to be relatively small since $\langle
\bv'_i \rangle = 0$ in our approximations. This is true also
in the mass and momentum equations.
Equations (\ref{mass2}), (\ref{momentum3}) and (\ref{energy_micro}) 
have the coarse grained form that will remain in the final 
DPD equations. Note, however, that they retain the full microscopic
information about the MD system, and for that reason they are 
time-reversible. Equation (\ref{momentum3}) for instance
contains only terms of even order in the velocity. In the next section
terms of odd order will appear when this equation is averaged.

It can be seen that the rate of change of momentum  in  \eq{momentum3})
is given as a sum of separate pairwise contributions from the other
particles, and that these terms  are all  odd under the 
exchange $l\leftrightarrow k$. Thus 
the particles interact in a pairwise fashion and
individually fulfill Newton's third law; in other words, 
momentum conservation is again explicitly upheld.
The same symmetries hold for the mass conservation 
equation (\ref{mass2}) and energy equation (\ref{aa}).

\section{Derivation of dissipative particle 
dynamics: average and fluctuating forces}

We can now investigate the average 
and fluctuating parts of Eqs.~(\ref{energy_micro}), (\ref{momentum3}) and 
(\ref{mass2}).
In so doing we shall need to draw on a hydrodynamic
description of the underlying molecular dynamics and 
 construct a statistical mechanical
description of our dissipative particle dynamics. 
For concreteness we shall take the hydrodynamic description
of the MD system in question to be that of a simple 
Newtonian fluid \cite{landau59}.
This is known to be a good description for MD
fluids based on Lennard-Jones or hard sphere potentials,
particularly in three dimensions~\cite{koplik95}.
Here we shall carry out the analysis for systems in two spatial 
dimensions; the generalization to three dimensions is straight forward,
the main difference being of a practical nature as the Voronoi 
construction becomes more involved.

We shall begin by specifying a scale separation
between the dissipative particles and the molecular dynamics 
particles by assuming that
\be
{|\bx_i - \bx_j|} << {|\br_k - \br_l|}  \; ,
\label{scale}
\ee
where $\bx_i$ and $\bx_j$ denote the positions of neighbouring MD particles. Such a scale separation is in
general necessary in order for the coarse-graining procedure to be
physically meaningful. Although for the most part in this paper we are
thinking of the molecular interactions as being mediated by short-range
forces such as those of Lennard-Jones type,
a local description of the interactions  will still be
valid for the case of long-range Coulomb interactions in an
electrostatically neutral system, provided that
the screening length is shorter than the width of the overlap region between
the dissipative particles. Indeed, as we shall show here,
the result of doing a local averaging is that the original Newtonian equations 
of
motion for the MD system become a set of Langevin equations 
for the dissipative particle dynamics. These Langevin equations admit
an associated Fokker-Planck equation.
 An associated fluctuation-dissipation relation 
relates the amplitude of the Langevin force
to the temperature and damping in the system.

\subsection{Definition of ensemble averages}

With the mesoscopic variables now available, 
we need to define the correct average corresponding to 
a dynamical state of the system.
Many MD configurations are consistent
with a  given value of the set $\{ \br_k, M_k, \bU_k, E_k  \}$, and 
averages are computed by means of an ensemble
of systems with common {\it instantaneous} values of the set $\{ \br_k, M_k, 
\bU_k, E_k  \}$.
This means that only the time derivatives of the set $\{ \br_k, M_k, \bU_k, E_k  
\}$,
i.e. the forces, have a fluctuating part. 
In the end of our development 
approximate distributions for $\bU_k$'s and $E_k$'s will
follow from the derived Fokker-Planck equations. These 
distributions refer to the larger equilibrium ensemble that contains all
fluctuations in $\{ \br_k, M_k, \bU_k, E_k  \}$.

It is  necessary, to compute the average  MD particle velocity 
$\langle \bv \rangle $ {\it between} dissipative particle
centers, given $\{ \br_k, M_k, \bU_k, E_k  \}$. This  velocity depends
on all neighboring dissipative particle velocities. However, for 
simplicity we shall only employ
a ``nearest neighbor'' approximation, which consists in assuming 
that $\langle \bv \rangle $ interpolates linearly between the 
two nearest dissipative particles.
This approximation is of the same nature as the approximation
used in the Newtonian fluid stress--strain relation which is linear
in the velocity gradient.
This implies that in the overlap region between dissipative particles 
$k$ and $l$
\be
\langle \bv' \rangle  = \langle \bv' \rangle (\bx ) = \frac{\bx' \cdot
\br_{kl}}{r_{kl}^{2}} \bU_{kl} \; ,
\label{average}
\ee
where the primes are defined  in Eqs.~(\ref{primes})
and $r_{kl}= |\br_k - \br_l|$.

A preliminary mathematical observation is useful in splitting
the equations of motion into average and fluctuating parts.
Let $r(\bx )$ be an arbitrary, slowly varying function on the 
$a^2/r_{kl}$ scale. Then 
we shall employ the approximation corresponding to a linear 
interpolation between DP centers, that 
$r(\bx ) =  (1/2) (r_k + r_l) $ where $\bx$ is a position
in the overlap region between DP k and l and $r_k$ and $r_l$
are values of the function $r$ associated with the DP centers k and l
respectively.
\begin{figure}
\centerline{\hbox{\psfig{figure=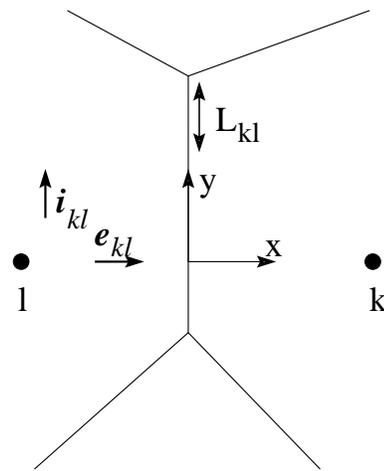,width=5cm}}}
\caption{\label{twist}
\protect \narrowtext 
Two interacting Voronoi cells. The length of the intersection
between DP's k and l is $l_{kl}$, the shift from the center
of the intersection between
$\br_{kl}$ and $l_{kl}$ is $L_{kl}$ ($L_{kl}=0$ when 
$\br_{kl}$ intersects $l_{kl}$ in the middle)
 and the unit vector $\bi_{kl}$ is normal to $\bee_{kl}$.
The coordinate system x-y used for the integration has its origin on
the intersection.}
\end{figure}
Then
 \bea 
& \sum_i & f_{kl}(\bx_i ) r(\bx ) \approx  \int dx \; dy  \frac{\rho_k + 
\rho_l}{2}  f_{kl}(\bx )
\frac{r_k + r_l}{2}  \nonumber \\
&\approx &   \frac{l_{kl}}{2a^2} \frac{\rho_k + \rho_l}{2}
\frac{r_k + r_l}{2} \int_{-\infty}^\infty dx'\; \cosh^{-2}{(x' r_{kl}/a^2 )}
\nonumber \\
&=&    \frac{l_{kl} }{r_{kl}} \frac{\rho_k + \rho_l}{2}\frac{r_k + r_l}{2},
\label{prefactor}
\eea
where $\frac{\rho_k + \rho_l}{2}$ is the MD particle number density 
and we have used the identity $\tanh'(x) 
= \cosh^{-2}(x)$.
We will also need the first moment in $\bx'$ 
\bea
\sum_i & f_{kl} (\bx_i ) & \bx'_i r(\bx_i )  \approx 
\int dx dy  \frac{\rho_k + \rho_l}{2}  f_{kl}(\bx ) \bx'
\frac{r_k + r_l}{2}\nonumber \\
& \approx & 
\frac{1}{2 a^2} \frac{\rho_k + \rho_l}{2} \frac{r_k + r_l}{2} 
\int dx \; dy   \cosh^{-2} 
\left(\frac{x r_{kl}}{ a^2} \right) y \bi_{kl}  \nonumber \\
&= & \frac{l_{kl}}{2 r_{kl}} L_{kl}  
 \frac{\rho_k + \rho_l}{2} \frac{r_k + r_l}{2} \bi_{kl}
\label{small}
\eea 
where the unit vectors $\bee_{kl} = \br_{kl}/r_{kl}$ and $\bi_{kl}$
are shown in Fig.~\ref{twist}, we have used the fact that the 
integral over $x \bee_{kl} \cosh^{-2} ...$ vanishes
since the integrand is odd,  and the last 
equation follows by the substitution $x \rightarrow (a^2/r_{kl}) x$.
In contrast to the vector $\bee_{kl}$
the vector $\bi_{kl}$ is even under the exchange $k \leftrightarrow l$,
as is $L_{kl}$. This is a matter of definition only as it would be equally
permissible to let $\bi_{kl}$ and $L_{kl}$ be odd under this exchange.
However, it is important for the symmetry properties of the 
fluxes that $\bi_{kl}$ and $L_{kl}$  have the same symmetry 
under $k \leftrightarrow l$.

\subsection{The mass conservation equation}

Taking the average of \eq{mass2}), we observe that   
the first  term vanishes if \eq{average}) is used,
and the second term follows directly from \eq{small}).
We thus obtain 
\be
\dot{M}_k = 
\sum_l (\la \dot{M}_{kl} \ra + \dm) 
\label{mass_dpd1}
\ee
where
\be  
 \langle \dot{M}_{kl}  \rangle = 
\sum_{li} f_{kl} m (\bx_i ) \la \bx'_i \ra \cdot \bU_{kl} =
\frac{l_{kl}}{2 r_{kl}} L_{kl}  \frac{\rho_k + \rho_l}{2}
\bi_{kl} \cdot \bU_{kl}
\label{mass0} \; ,
\ee
and $\dm =  \dot{M}_{kl} - \la \dot{M}_{kl} \ra $. 
The finite value of $\langle \dot{M}_{kl}  \rangle $ is caused
by the relative DP motion {\em perpendicular} to $\bee_{kl}$.
This is a geometric effect intrinsic to the Voronoi lattice.
When particles move the Voronoi boundaries change
their orientation, and this boundary twisting causes
mass to be transferred between DP's. This mass 
variation will be visible in  the energy flux, though not
in the momentum flux.
 It will later be shown that the effect of
mass fluctuations in the momentum and energy equations
may be absorbed in the force and heat flux fluctuations.

\subsection{The momentum conservation equation}

Using \eq{mass0}) we may split \eq{momentum3}) into 
average and fluctuating parts to get 
\bea
  \frac{\dd  \bP_k  }{\dd t} &=&   M_k  \bg \nonumber \\
  &+& \sum_l \langle \dot{M}_{kl} \rangle \frac{\bU_{k} + \bU_l}{2} 
+ \sum_{li}         f_{kl}(\bx_i )
\langle  {\bf \Pi}_i \rangle   \cdot \br_{kl}   \nonumber \\
&+&     \sum_i  f_{kl} (\bx_i ) m  \langle  \bv_i' \bx_i' \rangle
  \cdot  \bU_{kl}   + \sum_l \tilde{\bF}_{kl}
\label{momentum4} \; ,
\eea
where the fluctuating force or, equivalently, the momentum flux is
\bea
\tilde{\bF}_{kl}&=&  \sum_i  f_{kl}(\bx_i )  [({\bf \Pi}_i -\langle {\bf
  \Pi}_i  \rangle ) \cdot \br_{kl} \nonumber \\
&+& m (\bv'_i \bx'_i - \la \bv'_i \bx'_i \ra )\cdot \bU_{kl} ] \nonumber \\
& + &
\dm \frac{\bU_{k} + \bU_l}{2}  \; .
\eea
Note that by definition $\tilde{\bF}_{lk} = -\tilde{\bF}_{kl}$.
The fact that we have absorbed 
mass fluctuations with the fluctuations in $\tilde{\bF}_{kl}$
deserves a comment. 
 In general force fluctuations will
cause mass fluctuations, which in turn will couple back to 
cause momentum fluctuations. The time scale over which this will
happen is $t_{\eta}= r_{kl}^2/\eta$, where 
$\eta$ is the dynamic viscosity of the MD system.
This is the time it takes for 
a velocity perturbation to decay over a distance of $r_{kl}$.
Perturbations mediated by the pressure, i.e. sound waves, will
have a shorter time. In the sequel we shall need to make the
assumption that the forces are Markovian, and it is clear that 
this assumption may only be valid on time scales larger than $t_{\eta}$.
Since the time scale of a hydrodynamic perturbation of size $l$, say, is 
also given as $l^2/\eta$ this restriction implies  the scale separation
requirement $r_{kl}^2 << l^2$, consistent with the scale $r_{kl}$ being
mesoscopic.

Since $\langle  {\bf \Pi}_i \rangle$ is in general 
dissipative in nature, Eq.~(\ref{momentum4}) 
and its mass- and energy analogue will be referred to as DPD1. 
It is at the point of 
taking the average in Eq.~(\ref{momentum4}) that time reversibility is
lost. Note, however, that we do not claim to treat the
introduction of irreversibility into the problem in a mathematically 
rigorous way. This is a very difficult problem
in general which so far has only been realized by rigorous methods 
in the case of
some very simple dynamical systems with well defined ergodic 
properties~\cite{pvcop92,oppvc94,pvcrrh90}. We shall
instead use the constitutive relation for a Newtonian fluid which, as
noted earlier, is an emergent property of Lennard-Jones and hard sphere
MD systems, to give
Eq.~(\ref{momentum4})  a concrete content.
The momentum-flux tensor then has the following simple form
\be
\rho \langle {\bf \Pi }_i\rangle  = m \rho \bv \bv + 
{\bf I} p  - \eta (\nabla \bv + (\nabla \bv )^T)
\label{constitutive}
\ee
where  $p$ is the pressure and $\bv$ the average velocity 
of the MD fluid, $^T$ denotes the transpose and $\bf I$ is the
identity tensor~\cite{landau59}.  In the above equation 
we have for simplicity assumed that the bulk viscosity $\zeta = (2/d)\eta$ where 
$d$ is 
the space dimension 2.
The modifications to include an independent $\zeta$ are 
completely straight forward. 

Using the 
assumption of linear interpolation (\eq{average})), 
the advective term $\rho \bv \bv$
vanishes in the frame of reference of the overlap region since 
there $\bv' \approx 0$.
The velocity gradients in \eq{constitutive}) may be evaluated 
using \eq{average}); the result is 
\be
\nabla \bv + (\nabla \bv )^T = \frac{1}{r_{kl}} 
\left( \bee_{kl} \bU_{kl}  + \bU_{kl} \bee_{kl}  \right)
\label{gradient} \; .
\ee

%************ SURFACE INTEGRAL ********
Note further that  $\sum_l   l_{kl} $ is in fact a surface integral
over the DP surface. Consequently 
\be 
\sum_l   l_{kl} \bee_{kl} g_k = 0
\label{surface_integral}
\ee
for any function $g_k$ that does not depend on $l$.
In particular we have   $\sum_l   l_{kl}\bee_{kl} (p_k + p_l)/2  = - \sum_l   
l_{kl}\bee_{kl}
 p_{kl}/2  $, where $p_{kl} = p_k - p_l$.
%********
Combining Eqs.~(\ref{constitutive}), (\ref{prefactor})
and (\ref{gradient}), \eq{momentum4})
then takes the form
\bea
 &&   
\frac{\dd   \bP_k }{\dd t} = 
 M_k  \bg +  \sum_l \langle \dot{M}_{kl} \rangle \frac{\bU_{k} + \bU_l}{2}
\nonumber \\
  &-& \sum_l   l_{kl}  \left( 
\frac{p_{kl}}{2} \bee_{kl} +
\frac{\eta}{r_{kl}}  \left( \bU_{kl}  + (\bU_{kl}
  \cdot  \bee_{kl})  \bee_{kl}  \right)
\right)\nonumber \\
&+& \sum_l \tilde{\bF}_{kl}
\label{momentum5} \; ,
\eea
where we have assumed that the pressure $p$, as 
well as the average velocity, interpolates 
linearly between DP centers, and we have omitted the 
$ \la \bv'_i \bx'_i \ra \approx 0 $ term.
Note that all terms except the gravity term 
on the right hand side of \eq{momentum5}) are odd when
$k \leftrightarrow l$. This shows that Newton's third law
is unaffected by the approximations made and that momentum
conservation holds exactly. The same statements
can be made for the mass equation and the energy equation.
The pressure will eventually follow from an equation of state
of the form $p_k= p(E_k,V_k,M_k)$ where $V_k$ is the volume 
and $M_k$ is the mass of DP $k$.

\subsection{The energy conservation equation}

% For the present purposes we shall limit our attention to the 
% ensemble averaged value of $E_k$. This is sufficient
% to obtain $p_k$, which itself is an averaged quantity.
% In the same spirit of maximum simplicity we shall use
% the ideal gas equation of state
% \be
% p_k V_k = \frac{M_k}{m} k_BT = \frac{2}{3} E_k
% \label{eq_state} \; . 
% \ee
% This, however is not a necessary choice. Any 
% equation of state written in terms of the energy,
% volume and pressure would do.
% A smaller compressibility than that of the ideal gas may be
% desirable in order to approach the 
% limit of incompressible fluid flow more quickly.
Splitting Eq.~(\ref{energy_micro}) into an average and a fluctuating part gives 
\bea
\dot{E}_k &=& \sum_{li} f_{kl} (\bx_i)\left(  \langle \bJ_{\epsilon i}' \rangle
- \langle \Pi'_i \rangle \cdot \frac{\bU_{kl}}{2} \right)
\cdot \br_{kl} \nonumber \\
&+& \sum_{li} f_{kl} (\bx_i) \langle \epsilon'_i \bx'_i 
\rangle \cdot \bU_{kl} \nonumber \\
&+&  \sum_l \frac{1}{2} 
\langle \dot{M}_{kl} \rangle \left( \frac{\bU_{kl}}{2} \right)^2 \nonumber \\
&-& \sum_l \tilde{\bF}_{kl} \cdot \frac{\bU_{kl}}{2} + \tilde{q}_{kl}\; . 
\label{energy_dpd1} 
\eea
where we have   defined
\bea
\tilde{q}_{kl} &=& \sum_i f_{kl}(\bx_i) 
(\bJ'_{\epsilon i} - \langle \bJ'_{\epsilon i} \rangle )\cdot \br_{kl}
+ \frac{\dot{\tilde{M}}_{kl}}{2}  
\left( \frac{\bU_{kl}}{2} \right)^2 \nonumber \\
&+& \sum_i f_{kl}(\bx_i) [  ( \epsilon'_i  \bx'_i- 
\la \epsilon'_i  \bx'_i \ra )
\nonumber \\
&-&  m  \frac{\bU_{kl}}{2} \cdot  \bv'_i \bx'_i  ] \cdot \bU_{kl}  
\eea
i.e. the fluctuations in the heat flux also contains
the energy fluctuations caused by mass fluctuations.
This is like the momentum case.

Note that in taking the average  in \eq{energy_dpd1}) the
${\bf \Pi} \cdot \bU_{kl}$ product presents no problem as
$\bU_{kl}$ is kept fixed under this average. If we had 
averaged over different values of $\bU_{kl}$ the
product of velocities in ${\bf \Pi} \cdot \bU_{kl}$ would 
have caused difficulties.
Equation (\ref{energy_dpd1})  is the third component in the description at the
DPD1 level.

The average of the energy flux vector $\bJ_{\epsilon}$
is taken to have the general form \cite{landau59}
\be
\rho \langle  \bJ_{\epsilon} \rangle
  = \epsilon \bv + \sigma \cdot \bv - \lambda \nabla T
\label{constitutive2}
\ee
where $\sigma = {\bf \Pi} - \rho \bv \bv$ is the stress tensor, and $\lambda$ 
the thermal conductivity
and $T$ the local temperature.
Note that in \eq{energy_micro}) only $\bJ'_{\epsilon}$ appears.
Since $\bv' \approx {\bf 0}$ we have $\langle \bJ'_{\epsilon} \rangle =
\lambda \nabla T$.
Averaging of \eq{energy_dpd1}) gives
\bea
&&\dot{E}_{k}  = -\sum_l l_{lk}   \lambda \frac{T_{kl} }{r_{kl}} \nonumber \\
&-& \sum_l l_{lk}  \left( \frac{p_{k}+p_l}{2} \bee_{kl} - \frac{\eta}{r_{kl}}
(\bU_{kl} + (\bU_{kl}\cdot \bee_{kl})\bee_{kl}) \right)  \cdot
\frac{\bU_{kl}}{2}  \nonumber \\
 &+& \sum_l
\frac{1}{2} \langle \dot{M}_{kl} \rangle \left(\frac{\bU_{kl}}{2} \right)^2
+  \frac{l_{kl}}{4 r_{kl}} L_{kl} 
\bi_{kl} \cdot \bU_{kl}
\left( \frac{E_k}{V_k} + \frac{E_l}{V_l} \right) 
\nonumber \\
&-&  \sum_l \tilde{\bF}_{kl} \cdot \frac{\bU_{kl}}{2} + \tilde{q}_{kl}\; .
\label{energy_dpd}
\eea
where $T_{kl} = T_k - T_l$ is the temperature 
difference between DP's $k$ and $l$, and   we have 
used linear interpolation to write 
$\la \epsilon'_1 \ra = (1/2)(E_k/V_k + E_l/V_l)$.
The first term above describes the heat flux according to Fourier's law.
The next non-fluctuating terms, which are multiplied by $\bU_{kl}/2$
represent the (rate of) work done by the interparticle forces, and the 
$\tilde{\bF}_{kl}$ term represents the work done by the fluctuating force.

As has been pointed out by Avalos et al and Espanol\cite{avalos97,espanol97}
the work done by $\tilde{\bF}_{kl}$ has the effect that it increases the thermal 
motion of the DP's at the expense of a reduction in $E_k$. This
is the case here as well since the above $ \tilde{\bF}_{kl} \cdot \bU_{kl}$
term always has a positive average due to the positive correlation between the 
force
and the velocity increments.

Equation (\ref{energy_dpd}) is identical in form to the energy equation 
postulated 
by Avalos and Mackie  \cite{avalos97}, save for the fact that here the 
conservative 
force $ (({p_{k}+p_{l}})/{2} )\bee_{kl} \cdot {\bU_{kl}}/{2}  $ (which sums
to zero under $\sum_k$) is present.  
The pressure forces in the present case 
correspond to the conservative forces in conventional DPD--it will be 
observed that they are both derived from a potential. 
However, while the conservative force in conventional DPD must be thought to be 
carried by some field external to the particles, the pressure 
force in our model has its origin within the particles themselves.
There is also a small difference between the present form of Fourier's law
and the description of thermal conduction employed by Avalos and Mackie.
While the heat flux here is taken to be linear in differences in $T$, 
Avalos and Mackie use a flux linear in differences in $ (1/T)$. As both 
transport laws are approximations valid to lowest order in differences in $T$, 
they should be considered  equivalent.

With the internal energy variable at hand it is possible
to update the pressure and temperature $T$ of the DP's
provided an equation of state for  the underlying MD system is assumed,
and written in the form $P=P(E,V,m)$ and $T=T(E,V,m)$.
For an ideal gas these are the well known relations $PV=(2/d)E$ and
$k_BT=(2/d)mE$.
%Please check. Gianni

Note that we only need the average evolution of the pressure and temperature.
The fluctuations of $p$  are already contained in 
$\tilde{\bF}_{kl}$ and the effect of temperature fluctuations is
contained within $\tilde{q}_{kl}$.
% The Fokker-Planck equation 
% with the energy treated as a fluctuating variable as
% well as the fluctuation-dissipation relation that 
% relates the  magnitude of the heat flow fluctuations
% to the thermal conductivity has been worked out in 
% Refs.~\cite{avalos97,espanol97}.

At this point we may compare the forces arising in the present model to 
those used in conventional DPD. In conventional DPD
the forces are pairwise and act in a direction 
parallel to $\bee_{kl}$, with a conservative part 
that depends only on $r_{kl}$
and a dissipative part proportional to 
$(\bU_{kl}\cdot \bee_{kl})\bee_{kl}$~\cite{hoogerbrugge92,espanol95b,marsh98c}. 
The forces in our new version of
DPD are pairwise too. 
The analog of the conservative force, $ l_{kl} (p_{kl}/2) \bee_{kl}$, 
is central and its $\br$ dependence is given by the Voronoi 
lattice. When there is no overlap $l_{kl}$ between dissipative particles
their forces vanish. (A cut--off distance,
beyond which no physical interactions are permitted, 
was also present in the earlier versions of DPD--see, for example, 
Ref.~\cite{hoogerbrugge92}--where it was introduced 
to simplify the numerical treatment.)
Due to the existence of an overlap region in our model, the 
dissipative force has both a component parallel to $\bee_{kl}$
and a component parallel to the relative velocity $\bU_{kl}$.
However, due to the linear nature
of the stress--strain relation in the Newtonian MD
fluid studied here, 
 this force has the same simple linear 
velocity dependence that 
has been postulated  in the literature. 

The friction coefficient
is simply the viscosity $\eta$ of the underlying fluid times
the geometric ratio $l_{kl}/r_{kl}$.
As has been pointed out both in the context of DPD \cite{espanol95}
and elsewhere, the viscosity  is generally {\em not} 
proportional to a friction coefficient between the particles.
After all, conservative systems like MD 
are also described by a viscosity.
Generally the viscosity will be caused by the combined effect
of particle interaction  (dissipation, if any) and the momentum
transfer caused by particle motion. The latter 
contribution is proportional to the mean free path.
The fact that  the MD viscosity $\eta$, the DPD viscosity
and the friction coefficient are one and the same
therefore implies that the mean free path effectively vanishes.
This is consistent with the space filling nature of the particles.
See Sec.~\ref{low_visc} for a further discussion of the zero viscosity limit.

Note that constitutive relations like Eqs.~(\ref{constitutive}) 
and (\ref{constitutive2})
are usually regarded as components
of a top-down or macroscopic description of a fluid.
However, any bottom-up mesoscopic description
necessarily relies on the use of some kind of averaging procedure;
in the present context, these relations represent a 
natural and convenient
although by no means a necessary choice of average. 
The derivation of emergent constitutive relations is itself part of the
programme of non-equilibrium statistical mechanics (kinetic theory), 
which provides a link 
between the microscopic and the macroscopic levels. However, as noted
above, no general and rigorous procedure for deriving such relations has 
hitherto been realised; in the present theoretical treatment, such 
assumed constitutive relations are therefore a necessary 
input in the linking of the microscopic and mesoscopic levels.

\section{Statistical mechanics of dissipative particle dynamics}

In this section we discuss the statistical properties of the DP's
with the particular aim of obtaining the magnitudes of $\tilde{\bF}_{kl}$ and
$\tilde{q}_{kl}$.
We shall follow two distinct routes that lead to the same 
result for these 
quantities, one based on the conventional Fokker-Planck description
of DPD\cite{avalos97}, and one based on Landau's and Lifshitz's
fluctuating hydrodynamics \cite{landau59}.

It is not straightforward to obtain  a general statistical mechanical
description of the DP-system. The reason is that the DP's,
which exchange mass, momentum, energy and volume,
are not captured by any standard statistical ensemble.
For the  grand canonical ensemble, the system in question
is defined as the matter within a fixed volume, and
in the case of a the isobaric ensemble the 
particle number is fixed.
Neither of these requirements hold for a DP in general.

A system which exchanges mass, momentum, energy and volume
without any further restrictions will generally 
be ill-defined as it will lose its identity in the course of time.
The DP's of course remain well-defined by virtue of the coupling 
between the momentum and volume variables: The DP volumes are defined
by the positions of the DP-centers and the DP-momenta govern
the motion of the DP-centers. Hence the quantities that are exchanged
with the surroundings are not independent and the ensemble must be 
constructed accordingly.

However, for present purposes we shall leave aside the
interesting challenge of designing the statistical mechanical
properties of such an ensemble, and derive the magnitude
of $\tilde{\bF}_{kl}$ and $\tilde{q}_{kl}$ 
from two different approximations.
The approximations are both justifiable from the assumption
that  $\tilde{\bF}_{kl}$ and $\tilde{q}_{kl}$ have a negligible
correlation time. It follows that their properties
may be obtained from the DP behavior on such short time
scales that the DP-centers may be assumed fixed in space.
As a result, we may take 
either the DP volume or the system of MD-particles fixed for
the relevant duration of time.
Hence for the purpose of getting $\tilde{\bF}_{kl}$ and
$\tilde{q}_{kl}$  we may  use either the isobaric ensemble, applied to
the DP system, or the grand canonical ensemble, applied to the MD system.
We shall find the same results from either route.
The analysis of the DP system using the isobaric ensemble follows the
standard procedure using the Fokker-Planck equation,
and the result for the equilibrium distribution is only valid
in the short time limit.
The analysis of the MD system 
corresponding to the grand canonical ensemble could
be conducted along the similar lines.  However, it is also possible 
to obtain the magnitude of $\tilde{\bF}_{kl}$ and
$\tilde{q}_{kl}$ directly
from the theory of fluctuating hydrodynamics since 
this theory is derived from coarse-graining
the fluid onto a grid. The pertinent fluid velocity and stress
fields thus result from averages over {\em fixed volumes} associated
with the grid points: Since mass flows freely between these volumes
the appropriate ensemble is thus the grand canonical one.

\subsection{The isobaric ensemble}

We consider the system of $N_k \gg 1$ 
MD particles inside a given DP$_k$ at 
a given time, say all the MD particles with positions 
that satisfy $f_k(\bx_i ) 
>1/2$
at time $t_0$. At later times it will be possible to 
associate a certain volume 
per
particle with these particles, and by definition the system they form will
exchange volume and energy but not mass. 
We consider all the remaining DP's as a thermodynamic bath with which 
DP$_k$ is in equilibrium.
The system defined in this way will be described
by the Gibbs free energy and the isobaric ensemble.
Due to the diffusive spreading
of MD-particles, this system will only initially coincide
with the DP; during this transient time interval, however, we may treat
the DP's as systems of fixed mass and describe them by the
approximation $\la \dot{M}_{kl} \ra =0$.
The magnitudes of $\tilde{q}$ and $\tilde{\bF}$
follow in the form of fluctuation-dissipation relations
from the Fokker-Planck equivalent of our Langevin equations.
The mathematics involved in obtaining fluctuation-dissipation relations
is essentially well-known from the literature \cite{espanol95b}, and
our analysis parallels that of  Avalos and Mackie~\cite{avalos97}.
However, the fact that the conservative part of the conventional DP forces
is here replaced by the pressure and that the present DP's
have a variable volume makes a separate treatment enlightening.

The probability $\rho (V_k,\bP_k,E_k)$ of finding DP$_k$ with a volume $V_k$, 
momentum $\bP_k$ and 
internal energy $E_k$ is then proportional to $\exp (S_T/k_B)$
where $S_T$  is the entropy of all DP's  given that 
the values $(V_k,\bP_k,E_k)$ are known for DP$_k$\cite{reif65}.
If $S'$ denotes the entropy of the bath we can write $S_T$ as
\bea
S_T &=& S'(V_T-V_k, \bP_T - \bP_k, E_T - \frac{P_k^2}{2M_k} - E_k) + S_k 
\nonumber \\
&\approx & S'(V_T, \bP_T, E_T ) - \frac{\partial S'}{\partial E} \left( 
E_k + \frac{P_k^2}{2M_k} \right) -\frac{\partial S'}{\partial V} V_k \nonumber 
\\
&-& \frac{\partial S'}{\partial \bP} \bP_k + S_k  
\eea
where the derivatives are evaluated at  $(V_T,\bP_T,E_T)$ 
and thus characterize the bath only.  
Assuming  that $\bP_T$ vanishes  there is nothing in the system 
to give the vector ${\partial S'}/{\partial \bP}$ a direction, and it must 
therefore vanish as well \cite{landau59c}.
The other derivatives give the pressure $p_0$ and temperature $T_0$
of the bath and we obtain
\be
S_T = S'(V_T, \bP_T, E_T ) - \frac{1}{T_0} \left( G_k + \frac{P_k^2}{2M_k} 
\right)
\ee
where the Gibbs free energy has the standard  form $G_k = E_k + p_0 V_k - T_0 
S_k$.
Since there is nothing special about DP$_k$ it immediately follows that the 
the full equilibrium distribution has the form
\be
\rho^{\text{eq}} = Z^{-1}(T_0,p_0)\exp \left( -\beta_0 \sum_k \frac{P_k^2}{2M_k} 
+ G_k \right) \; ,
\label{distribution}
\ee
where $\beta_0 = 1/(k_BT_0)$.
The temperature $T_k = (\partial S_k /\partial E_k)^{-1} $ and pressure 
$p_k =   T_k (\partial S_k /\partial V_k) $ will fluctuate around the
equilibrium values $T_0$ and $p_0$.
The above distribution is analyzed by Landau and Lifshitz \cite{landau59c} who
show that the fluctuations have the magnitude
\be
\langle \Delta P_k^2\rangle  = \frac{k_BT_0}{V_k \kappa_S }, 
\; \;  \langle \Delta T^2_k \rangle  = \frac{k_BT_0^2}{V c_v } 
\label{fluctuations}
\ee
where the isentropic compressibility $\kappa_S = -(1/V) (\partial V/\partial 
P)_S$ and
the specific heat capacity $c_v$ are both intensive quantities.
Comparing our expression 
with the distribution postulated by Avalos and Mackie, we have 
replaced the Helmholtz by the Gibbs free energy in \eq{distribution}).
This is due to the fact that our DP's exchange volume as well as 
energy.

We write the fluctuating force as
\be
\tilde{\bF}_{kl} = \bom_{kl \parallel} W_{kl\parallel}  + \bom_{kl \perp} 
W_{kl\perp}
\label{fluc_force}
\ee 
where, for reasons soon to become apparent, we have chosen
to decompose $\tilde{\bF}_{kl}$ into components parallel
and perpendicular to $\bee_{kl}$.
The $W$'s are defined as Gaussian random variables with the
 correlation function 
\bea
\langle  W_{kl \alpha} (t) W_{nm \beta}(t') \rangle 
&=& \delta_{\alpha \beta} \delta (t-t') (\delta_{kn}\delta_{lm}
+\delta_{km}\delta_{ln})
 \label{correlations}
\eea
where $\alpha$ and $\beta$ denote either $\perp$ or $\parallel$.
The product of $\delta$ factors ensures that only 
equal vectorial components of the forces between a pair
of DP's are correlated, while Newton's third law  
guarantees that $\bom_{kl} = - \bom_{lk}$. 
Likewise the fluctuating heat flux takes the form
\be
\tilde{q}_{kl} = \Lambda_{kl} W_{kl}
\ee
where $W_{kl}$ satisfies \eq{correlations}) without the 
$\delta_{\alpha \beta}$ factor and energy conservation implies
$\Lambda_{kl}  = -\Lambda_{lk} $.

The force correlation function then takes the form 
\bea 
\langle   \tilde{\bF}_{kn}(t) \tilde{\bF}_{lm}(t') \rangle  &=& 
(\bom_{kn\perp}  \bom_{lm\perp} + \bom_{kn\parallel}  \bom_{lm\parallel})
\nonumber    \\ 
&& ( \delta_{kl} \delta_{nm}
+ \delta_{km} \delta_{ln} )  \delta (t-t') \nonumber \\
&\equiv & \bom_{klnm} ( \delta_{kl} \delta_{nm}
+ \delta_{km} \delta_{ln} )  \delta (t-t') \;
\label{force_correlations}
\eea
where we have introduced the second order tensor $ \bom_{knlm}$.

It is a standard result in non-equilibrium statistical mechanics 
that a Langevin
description of a dynamical variable $\bf y$
\be 
\dot{\bf y} =   {\bf a} ({\bf y}) + \tilde{\bf G}
\label{langevin}
\ee
where $\tilde{\bG}$ is a delta-correlated force
has an equivalent probabilistic representation 
in terms of the Fokker-Planck equation
\be 
\frac{\partial \rho({\bf y},t)}{\partial t} = -\nabla \cdot 
 ({\bf a} ({\bf y}) \rho({\bf y})) + \frac{1}{2} \nabla \nabla  
\colon ({\bf A}({\bf y})
 \rho({\bf y}))
\ee
where $\nabla $ denotes derivatives with respect to ${\bf y}$
and  $\rho({\bf y},t)$ is the probability
distribution for the variable ${\bf y}$ at time $t$, 
 $ \langle 
 \tilde{\bG}({\bf y},t) \tilde{\bG}({\bf y},t')\rangle   = {\bf A} \delta (t-
t')$ and
$\bf A$ is a symmetric tensor of rank two~\cite{gardiner85}.

In the preceding paragraph, $\bG$ denotes all the fluctuating terms in 
Eqs.~(\ref{momentum5})
and (\ref{energy_dpd}). 
Using the above definitions and $\la \dot{M}_{kl} \ra = 0$
it is a standard matter \cite{espanol95b} to obtain the Fokker-Planck
equation
\be
\frac{\partial \rho}{\partial t} = (L_0 + L_{\text{DIS}} +L_{\text{DIF}}), \rho
\label{fokker_planck}
\ee
where 
\bea
L_0 &=& -\sum_k \frac{\partial }{\partial \br_k} \cdot \bU_k +
\sum_{k\neq l} l_{kl} \left( \frac{\partial }{\partial \bP_k} \cdot \bee_{kl} 
\frac{p_{kl}}{2} \right. \nonumber \\
&+& \left.   \frac{\partial }{\partial E_k } \bee_{kl} \cdot \bU_{kl} \frac{p_k 
+ p_l}{4} \right) \nonumber \\
 L_{\text{DIS}} &=&  \sum_{k\neq l} l_{kl} \left( \frac{\partial }{\partial 
\bP_k } \cdot
\bF^D_{kl} -     \frac{\partial }{\partial E_k }\left( \frac{\bU_{kl}}{2} \cdot  
\bF^D_{kl}  -
\lambda \frac{T_{kl}}{r_{kl}}  \right) \right) \nonumber \\
 L_{\text{DIF}} &=& \frac{1}{2} \sum_{k\neq l}  \left(  \bom_{klkl} \cdot 
\frac{\partial }{\partial \bP_k } \cdot
\bL_{kl} -    \frac{\partial }{\partial E_k } \left( \bom_{klkl} \cdot 
\frac{\bU_{kl}}{2}  \cdot \bL_{kl}  \right. \right.  \nonumber \\
&-& \left. \left.  \Lambda^2_{kl} \left(   \frac{\partial }{\partial E_k } -  
\frac{\partial }{\partial E_l }
\right) \right) \right),
\label{operators}
\eea
$\bF^D_{kl} = (\eta/r_{kl}) (\bU_{kl} + (\bU_{kl}\cdot \bee_{kl} ) 
\bee_{kl} )$, and
the sum $\sum_{k\neq l}$ runs over both $k$ and $l$. The operator $\bL_{kl}$
is defined as in Ref.~\cite{avalos97}:
\be
\bL_{kl}  = \left(  \frac{\partial }{\partial \bP_k } 
-  \frac{\partial }{\partial \bP_l }\right) - \frac{\bU_{kl}}{2} 
\left(  \frac{\partial }{\partial E_k } -  \frac{\partial }{\partial E_l } 
\right) \; .
\ee

The steady-state solution of \eq{fokker_planck}) is already given by 
\eq{distribution}); following conventional procedures we can
obtain the fluctuation-dissipation relations
for $\bom$ and $\Lambda$ by inserting  $\rho^{\text{eq}}$ in 
\eq{fokker_planck}).

Apart from the tensorial nature of $\bom_{klkl}$ the  operators $ 
L_{\text{DIS}}$
and $ L_{\text{DIF}}$ are essentially identical to those published
earlier in conventional DPD~\cite{avalos97,espanol97}.
However, the `Liouville' operator $L_0$ plays a somewhat different role 
as it contains the $\partial /\partial E_k$ term, corresponding to the fact that 
the pressure forces do work on the DP's to change their internal energy.

While $L_0 \rho^{\text{eq}}$  conventionally vanishes exactly
by construction of the inter-DP forces, here it vanishes 
only to order $1/N_k$. 
In order to evaluate  $L_0 \rho^{\text{eq}}$   we need the following 
relationship
\be 
 \frac{\partial }{\partial \br_k } = \frac{1}{2} \sum_{k\neq l} l_{kl} \bee_{kl}
 \left(  \frac{\partial }{\partial V_l }  -  \frac{\partial }{\partial V_k } 
\right) \; ,
\ee
which is derived by direct geometrical consideration of the
 Voronoi construction. By repeated use of 
\eq{surface_integral}) it is then a straightforward algebraic task 
to obtain 
\be
L_0 \rho^{\text{eq}} = \frac{\rho^{\text{eq}} }{4} \sum_{k\neq l} l_{kl} 
\bee_{kl}\cdot \bU_k  \left[  \frac{\partial p_l}{\partial E_l} -\frac{ p_{kl} 
T_{kl}}{k_BT_kT_l}  \right] \; ,
\label{L0}
\ee
which does not vanish identically.
However, note that if we estimate $E_l \approx N_l k_BT$ we obtain ${\partial 
p_l}/{\partial E_l} \approx (1/N_k) (p_l/k_BT)$.
Similarly we may estimate $p_{kl} $ and $T_{kl}$ from \eq{fluctuations}) to 
obtain
\be
\frac{p_{kl}T_{kl}}{k_BT_kT_l} \approx \frac{\sqrt{ \Delta P^2 \Delta 
T^2}}{k_BT_kT_l} = \frac{1}{N_k}
\sqrt{\frac{N_k/V_k}{\kappa_S c_v T_0^2}} \; .
\ee
The last square root is an intensive quantity of the order $p_0/(k_BT_0)$, as 
may be easily demonstrated for the case
of an ideal gas.
Since each separate quantity that is contained in the differences in the  square 
brackets of \eq{L0}) is of the 
order $p_0/T_0$ we have shown that they cancel up to relative order $1/N_k\ll 1$.
In fact, it is not surprising that Langevin equations which approximate 
local gradients to first order only in the corresponding differences, 
like $T_{kl}$, give rise to a Fokker-Planck description 
that contains higher order correction terms.

Having shown that $L_0 \rho^{\text{eq}}$ vanishes to a good approximation 
we may proceed to obtain the fluctuation-dissipation relations from
the equation $ ( L_{\text{DIS}}  +  L_{\text{DIF}})  \rho^{\text{eq}} = 0$.
It may be noted from \eq{operators}) that this equation is satisfied if 
\bea
(l_{kl} \bF^D_{kl} + \frac{1}{2} \bom_{klkl} \bL_{kl} )  \rho^{\text{eq}} &=& 0 
\nonumber \\
\left( l_{kl} \lambda \frac{T_{kl}}{r_{kl}} + \frac{1}{2} \Lambda^2_{kl} \left( 
\frac{\partial }{\partial E_k } -  \frac{\partial }{\partial E_l }\right)\right) 
\rho^{\text{eq}}  &=& 0\; .
\label{fd0}
\eea
Using the identity
\be
\bee_{kl}\bee_{kl} + \bi_{kl}\bi_{kl} = {\bf I}
\ee
where $\bi_{kl}$ a 
vector normal to $\bee_{kl}$, 
we may show that \eq{fd0}) implies that 
\bea
\omega_{kl\parallel}^2 &=& 2 \omega_{kl\perp}^2 
=4 \eta k_B \Theta_{kl} \frac{l_{kl} }{ r_{kl} } \nonumber \\
\Lambda_{kl}^2 &=& 2 k_B T_k T_l \lambda \frac{l_{kl}}{r_{kl}} \; ,
\label{fd} 
\eea
where $\Theta^{-1}_{kl} = (1/2) ( T_k^{-1} + T_l^{-1} ) $.

%******END PAD

\subsection{$\tilde{\bF}$ from fluctuating hydrodynamics}

Having derived  the fluctuation-dissipation relations from the 
approximation of the isobaric ensemble we now derive the 
same result from fluctuating hydrodynamics, which corresponds 
to the grand canonical ensemble. We shall only derive the 
magnitude of $\tilde{\bF}_{kl}$ since $\tilde{q}$ follows 
on the basis of the same
reasoning.

Fluctuating hydrodynamics \cite{landau59} is based on the conservation equations
for mass, momentum and energy with the modification that the momentum 
and energy fluxes contain an additional fluctuating term.
Specifically, the momentum flux tensor takes the form
$-\nabla P + \rho \bv \bv + \sigma'$, where $P$ is the pressure,
$\bv$ is the velocity field and the viscous stress tensor is given as
\be 
\sigma' = \eta \left( \nabla \bv + \nabla \bv^T - \frac{2}{d}\nabla \cdot \bv 
\right)
+ \zeta \nabla \cdot \bv + \bs ,
\ee
where $\bs$ is the fluctuating component of the momentum flux.
{}From the same approximations as we used in deriving \eq{fd}), i.e.
a negligible correlation time for the fluctuating forces,
Landau and Lifshitz derive
\bea
&& \la \bs (\bx , t) \cdot \bn  \bs (\bx' , 0)\cdot \bn   \ra = 2 k_BT \left( 
\eta (1+ \bn \bn )+ (\zeta - \frac{2}{d} \eta ) \bn \bn \right)  \nonumber \\
&& \delta (t) \left\{ \begin{array}{cc}
\frac{1}{\Delta V_n} & \mbox{if } \bx , \bx' \varepsilon \Delta V_n \\
0 &  \mbox{otherwise}
\end{array} \right .  \; 
\eea
where $\bn$ is an arbitrary unit vector and $n$ labels the volume element 
$\Delta V_n$.
By following the derivations presented by Landau and Lifshitz, 
it may be noted that nowhere is it
assumed that
 the $\Delta V_n$'s are cubic or stationary.
\begin{figure}
\centerline{\hbox{\psfig{figure=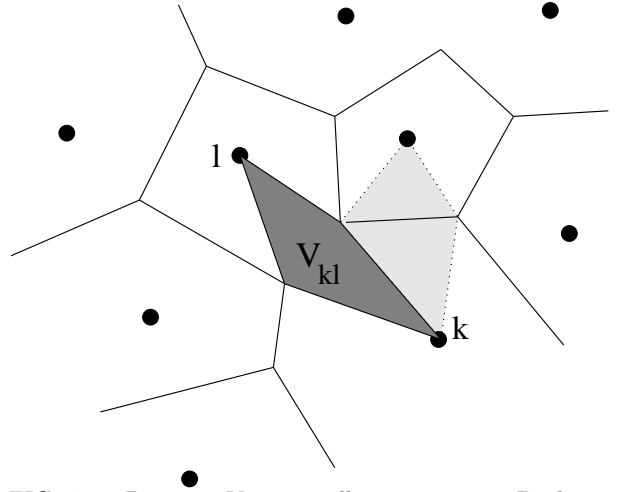,width=8cm}}}
\caption{\label{figV}
\protect \narrowtext 
Pairwise Voronoi-cell interactions. Dark gray: The volume $V_{kl}$ associated with the interaction 
between a single DP-pair. The light gray region shows the 
volume of the neighboring interaction.}
\end{figure}
By making the identifications $\zeta = (2/d) \eta $, $\bn \rightarrow \bee_{kl} 
$  $\tilde{\bF}_{kl} = 
l_{kl} \bs \cdot \bee_{kl} $, $\Delta V_n \rightarrow V_{kl}$,
(shown in Fig.~\ref{figV}), and 
$T = \Theta_{kl}$ we may immediately write down 
\bea
\la \tilde{\bF}_{kl} (t) \tilde{\bF}_{nm} (0) \ra &= &
2 \frac{k_B \Theta_{kl}l_{kl}^2}{V_{kl}} \eta
(1 + \bee_{kl}  \bee_{nm} ) \delta (t) \nonumber \\
& (& \delta_{kn}\delta_{lm} + 
\delta_{km}\delta_{ln} )
\label{grunch}
\eea 
where again the last 
sum of $\delta$-factors ensures that $kl$ and $nm$ denote the same DP 
pair.
Observing from Fig.~\ref{figV} that 
$V_{kl}= l_{kl} r_{kl}$, 
it now follows directly from \eq{grunch}) that 
\bea
\la \tilde{\bF}_{kl} (t) \cdot \bee_{kl} \tilde{\bF}_{nm} (0) \cdot \bee_{nm}  
\ra
&=& 2 \la \tilde{\bF}_{kl} (t) \cdot \bi_{kl} \tilde{\bF}_{nm} (0) \cdot 
\bi_{nm}  \ra \nonumber \\
&=& 4 k_B \Theta_{kl} \frac{l_{kl}}{r_{kl}}
 \eta \delta (t) \nonumber \\
&(& \delta_{kn}\delta_{lm} + \delta_{km}\delta_{ln} )
\label{fd2}
\eea
which is nothing but the momentum part of \eq{fd}). That the
fluctuating heat flux
$\tilde{q}$ produces the form of 
fluctuation-dissipation relations given in \eq{fd}) 
follows from a similar analysis.
Thus the approximation of fixed DP volume 
$V_k$ produces the
same result as the approximation of fixed number of MD particles
$N_k$. This is due to the fact that both approximations 
are based on the assumption
that the DP's are only considered within a time interval 
which is longer than the 
correlation time of the fluctuations but shorter than the time 
needed for the DP's to move significantly.

The result given in \eq{fd2}) was derived from the 
somewhat arbitrary choice of discretizaton volume $V_{kl}$;
this is the volume which corresponds to the segment $l_{kl}$
over which all forces have been taken as constant. It is thus 
the smallest discretization volume we may consistently choose.
It is reassuring that \eq{fd2}) also follows from different
choices of $\Delta V_n$. For example, one may readily 
check that \eq{fd2}) is obtained
if we split $V_{kl}$ in two along $r_{kl}$ and 
consider $\tilde{\bF}_{kl}$ to be the sum of two independent
forces acting on the two parts of $l_{kl}$.

We are now in a position to quantify the average component 
$\langle \dot{\tilde{E}}_k \rangle \equiv \sum_{l\neq k} \langle 
\tilde{\bF}_{kl} \cdot \bU_{kl}/2\rangle $ 
of the fluctuations in the internal energy 
given in \eq{energy_dpd}). 
Writing the velocity in response to $\tilde{\bF}_{kl}$ as 
$\tilde{\bU}_k = \sum_{l\neq k} \int_{-\infty}^t \dd t' \tilde{F}_{kl}(t')/M_k$,
we get that $ \langle \dot{\tilde{E}}_k \rangle = \sum \int_{-\infty}^t \dd t' 
\langle 
\tilde{F}_{kl}(t') \tilde{F}_{kl}(t) \rangle$ which by 
Eqs.~(\ref{fd}) and (\ref{force_correlations}) becomes
$\langle \dot{\tilde{E}}_k \rangle = (1/M_k) \sum 3 l_{kl} \eta k_B 
\Theta_{kl}/r_{kl} $.
This result is the same as one would have obtained applying the rules of
It\^{o} calculus to $\tilde{\bU}_k^2/(2M_k)$. 
It yields the modified, though equivalent, energy equation
\bea
&&\dot{E}_{k}  = -\sum_l l_{lk}   \lambda \frac{T_{kl} }{r_{kl}} \nonumber \\
&-& \sum_l l_{lk}  \left( \frac{p_{k}+p_l}{2} \bee_{kl} - \frac{\eta}{r_{kl}}
(\bU_{kl} + (\bU_{kl}\cdot \bee_{kl})\bee_{kl}) \right)  \cdot
\frac{\bU_{kl}}{2}  \nonumber \\
&-&  \sum_l \tilde{\bF'}_{kl} \cdot \frac{\bU_{kl}}{2} - 
3 \frac{l_{kl}}{r_{kl}} \eta k_B \Theta_{kl} + \tilde{q}_{kl}\; .
\label{energy_dpd2}
\eea
where we have written $\tilde{\bF'}_{kl}$ with a prime to denote
that it is uncorrelated with $\bU_{kl}$. In a numerical implementation  
this implies that $\tilde{\bF'}_{kl}$  must be 
generated from a different random variable than $\tilde{\bF}_{kl}$,
which was used to update $\bU_{kl}$.

The fluctuation-dissipation relations Eqs.~(\ref{fd})
 complete our theoretical description of dissipative
particle dynamics, which has been derived by a coarse-graining of
molecular dynamics. All the parameters and properties of this new
version of DPD are related directly to  the underlying molecular
 dynamics, and properties such as the viscosity which are 
emergent from it.

\section{Simulations}

While the present paper primarily deals with theoretical
developments we have carried out simulations to test the 
equilibrium behavior of the model in the case of the isothermal model.
 This is a crucial test 
as the derivation of the fluctuating forces relies on the
most significant approximations.
The simulations are carried out using a periodic Voronoi tesselation
described in detail elsewhere~\cite{defabritiis99}.

\begin{figure}
\centerline{\hbox{\psfig{figure=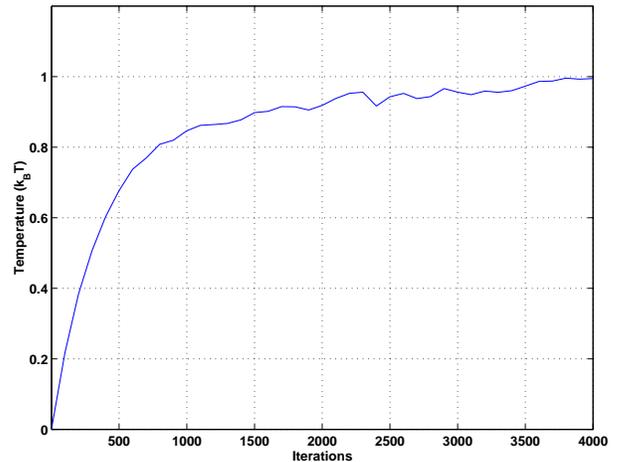,width=8cm}}}
\caption{\label{figkT}
\protect \narrowtext 
The DPD temperature (energy units) averaged over 5000 dissipative
particles as a function of time 
(iteration number in the integration scheme), 
showing good convergence to the underlying
molecular dynamics temperature which was set at one. This simulation
provides strong support for the approximations used to derive the
fluctuation-dissipation relations 
in our DPD model from molecular dynamics.}
\end{figure}

Figure \ref{figkT} shows the relaxation process towards equilibrium of an 
initially
motionless system. The DP temperature is measured as
$\langle \bP_k^2/(2 M_k) \rangle$ for a system of DPs 
with internal energy equal to unity.
The simulations were run for 4000 iterations of 5000 dissipative
particles and a timestep
$\dd t =0.0005$ using an initial molecular density $\rho =5$ for each DP.
The molecular mass was taken to be $m=1$, 
the viscosity was set at $\eta=1$, 
the expected mean free path is 0.79, 
and the Reynolds number (See Sec.~\ref{low_visc}) is Re=2.23.
It is seen that the convergence of the DP system 
towards the MD temperature is good, a result that provides
strong support for the 
fluctuation-dissipation relations of \eq{fd}).

\section{Possible applications}
\subsection{Multiscale phenomena}
 
For most practical applications involving complex fluids, additional 
interactions and boundary conditions need to be specified.
These too must be 
deduced from the microscopic dynamics, just as we have done for 
the interparticle forces.
This may be achieved by considering a 
particulate
description of the boundary itself and including molecular interactions
between the fluid MD particles and other objects, such as 
particles or walls. Appropriate modifications can then be made
on the basis of the
momentum-flux tensor of \eq{momentum_flux}), which 
is  generally valid.

\begin{figure}
\centerline{\hbox{\psfig{figure=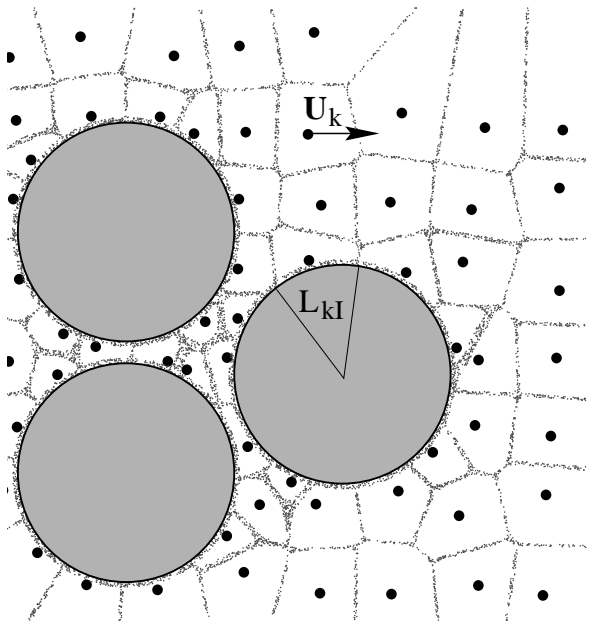,width=8cm}}}
\caption{\label{fig3}
\protect \narrowtext  {\bf Multiscale modeling of colloidal fluids.}
As usual, the dissipative particles are defined as cells
in the Voronoi lattice. Note that there are four relevant length
scales in this problem: the scale of the large, gray colloid particles,
the two distinct scales of the dissipative particles in between and
away from the colloids and finally the molecular scale of the 
MD particles. These mediate the  mesoscopic 
interactions and are  shown as dots on the 
boundaries between dissipative and 
colloidal particles.}
\end{figure}

Consider for example the case of a colloidal suspension, which is
shown in Fig.~\ref{fig3}. Beginning with 
the hydrodynamic momentum-flux tensor \eq{momentum_flux}) and 
\eq{momentum5}), it is evident that we also need to define 
an interaction region where the DP--colloid forces act:
the DP--colloid interaction may be 
obtained in the same form as the DP--DP interaction 
of \eq{momentum5}) by making the replacement $l_{kl}\rightarrow L_{kI}$,
where  $L_{kI}$  is the length (or area in 3D) of the arc 
segment where the dissipative particle
meets the colloid (see Fig.~\ref{fig3}) and the velocity 
gradient 
$r_{kl}^{-1}(  ({\bf U}_{kl}\cdot  {\bf e}_{kl}) {\bf e}_{kl} 
   + {\bf U}_{kl})$ 
is that between the dissipative particle 
and the colloid surface. The latter may be computed 
using ${\bf U}_k $ and the velocity of the colloid surface 
together with a no-slip boundary condition on this surface.
In \eq{fd}) the replacement $l_{kl} \rightarrow L_{KI}$
must also be made. 

Although previous DPD simulations of colloidal fluids have proved
rather successful~\cite{boek97} at low to intermediate solids 
volume fractions, they break down for dense systems
whose solids volume fraction exceeds a value of about 
40\% because the existing
method is unable to handle multiple lengthscale phenomena. However,
our new version of the algorithm provides the freedom to define
dissipative particle sizes according to the local resolution
requirements as illustrated in Fig.~\ref{fig3}. 
In order to increase the spatial resolution where 
colloidal particles are within close proximity 
it is necessary and perfectly admissible to introduce a 
higher density of dissipative particles
there; this ensures that fluid lubrication and hydrodynamic 
effects are properly maintained.
After these dissipative particles have moved it may be 
necessary to re-tile
the DP system; this is easily achieved by distributing 
the mass and momentum of the old dissipative particles on 
the new ones according to 
their area (or volume in 3D). Considerations of space prevent us from
discussing this problem further in the present paper, but 
we plan to report in detail on such
dense colloidal particle simulations using our method in 
future publications. We note in passing that a wide variety of other
complex systems exist where modeling and simulation are challenged by
the presence of several simultaneous length scales, for example in
polymeric and amphiphilic fluids, particularly in confined geometries
such as porous media~\cite{coveney98}.

\subsection{The low viscosity limit and high Reynolds numbers}
\label{low_visc}
In the kinetic theory derived by Marsh, Backx and Ernst [15]
% \cite{MBE1}
the viscosity is explicitly shown to have a kinetic
contribution  $\eta_K = \rho D/2$ where $D$ is  the DP self diffusion 
coefficient 
and $\rho$ the mass density.
The kinetic contribution to the viscosity was measured by Masters and  Warren 
\cite{masters99}
within the context of an improved theory.
How then can the viscosity $\eta$ used in our model 
be decreased to zero while kinetic 
theory
puts the lower limit $\eta_K$ to it?

%\subsubsection{Non-dimensionalizing the DPD equations}

To answer this question we must define a physical  way of decreasing the MD 
viscosity
while keeping other quantities fixed, or,  alternatively rescale the system in a 
way that 
has the equivalent  effect.
The latter method is preferable as it allows the underlying microscopic system 
to 
remain fixed.
 In order to do this we non-dimensionalize the DP
momentum equation \eq{momentum5}).

For this purpose we introduce the characteristic equilibrium velocity, $U_0 = 
\sqrt{k_BT/M}$,
the characteristic distance $r_0$ as the typical DP size. Then   the 
characteristic
time $t' = r_0/U_0$ follows. 
% \eq{momentum5}) 

Neglecting gravity for the time being \eq{momentum5}) 
 takes the form
\bea
 &&    \frac{\dd   \bP_k' }{\dd t'} = 
  - \sum_l   l'_{kl}  \left( 
\frac{p'_{kl}}{2} \bee_{kl} +\frac{1}{\text{Re}}  \left( \bU'_{kl}  + (\bU'_{kl}
  \cdot  \bee_{kl})  \bee_{kl}  \right)
\right)\nonumber \\
&+& 
\sum_l \frac{l'_{kl}L'_{kl}}{2 r'_{kl}} \frac{\rho'_k  + \rho'_l }{2}
\bi_{kl}\cdot \bU'_{kl}\frac{\bU'_k + \bU'_l}{2}
+ \sum_l \tilde{\bF}'_{kl}
\label{non_dim} \; ,
\eea
where $\bP_k'= \bP_k/(MU_0)$, $p'_{kl}= p_{kl}r_0^2/(MU_0^2)$,
$M = \rho r_0^2$ in 2d, the Reynolds number $\text{Re} = U_0 r_0\rho /\eta$
and  $\tilde{\bF}'_{kl}= (r_0/MU_0^2)\tilde{\bF}_{kl}$ 
where $\tilde{\bF}_{kl}$ is given by Eqs.~(\ref{fluc_force}) and (\ref{fd}).
A small calculations then shows that if  $\tilde{\bF}'_{kl}$ is 
related to $ \omega_{kl}'$ and $t'$ like $\tilde{\bF}_{kl}$ related to 
$ \omega_{kl}$ and $t$, then
\be
\omega_{kl}^{'2} \approx \frac{1}{\text{Re}} \frac{k_BT}{M U_0^2} \approx 
\frac{1}{\text{Re}} 
\ee
where we have neglected  dimensionless geometric  prefactors like 
$l_{kl}/r_{kl}$ and used
the fact that the ratio of the thermal to kinetic energy by definition of $U_0$ 
is one.

The above results imply that when the DPD system 
is measured in non-dimensionalized 
units 
everything is determined by the value of the mesoscopic Reynolds number Re.
There is thus no observable difference in this system  between increasing 
$r_0$ and decreasing $\eta$.

Returning to dimensional units again the DP diffusivity may be obtained  from 
the 
Stokes-Einstein relation \cite{einstein05} as 
\be
D= \frac{k_BT}{a r_0 \eta}
\label{einstein}
\ee
where $a$ is some geometric factor ($a=6\pi $ for a sphere) and 
all quantities on the right hand side except $r_0$ refer directly to the
underlying MD. 
As we are keeping the MD system fixed and 
increasing Re by increasing  $r_0$, 
it is seen that $D$ and hence $\eta_K$ vanish in the process.

We note in passing that if $D$ is written in terms of the mean free path 
$\lambda$:
$D=\lambda \sqrt{k_BT/(\rho r_0^2)}$ and this result is compared with 
\eq{einstein})
we get $\lambda' = \lambda /r_0 \sim 1/r_0 $ in 2d, i.e. the mean free path, 
measured
in units of the particle size decreases as the inverse particle size.
This is consistent with the decay of $\eta_K$.
The above argument shows that decreasing $\eta$ is equivalent to 
keeping the microscopic MD system fixed while increasing the DP size,
in which case the mean free path effects on viscosity is decreased to zero
as the DP size is increased to infinity. It is in this limit that 
high Re values may be achieved.

Note that in this limit the thermal forces $\tilde{\bF}_{kl} \sim \text{Re}^{-
1/2}$
will vanish, and that we are effectively left with  a macroscopic, 
fluctuationless
description.
 This is no problem when using the present Voronoi construction.
However, the effectively spherical particles of conventional DPD will 
freeze into a colloidal crystal, i.e. into a 
lattice configuration [8,9] in this 
limit.
Also while conventional DPD has usually 
required calibration simulations
to determine the viscosity, due to discrepancies between theory
and measurements, 
the viscosity in this new form of DPD is simply an input parameter.
However, there may still be  
discrepancies due to the approximations made in going from 
MD to DPD. These approximations include the linearization of the  
inter-DP velocity fields,
the Markovian assumption in the force correlations 
and the neglect of a DP angular momentum variable.

None of the conclusions from the above arguments would change if 
we had worked in three dimensions in stead of two.

\section{Conclusions}

We have introduced a systematic procedure 
for deriving the mesoscopic modeling and simulation method known as 
dissipative particle dynamics 
from the underlying description in terms of molecular dynamics. 
\begin{figure} 
\centerline{\hbox{\psfig{figure=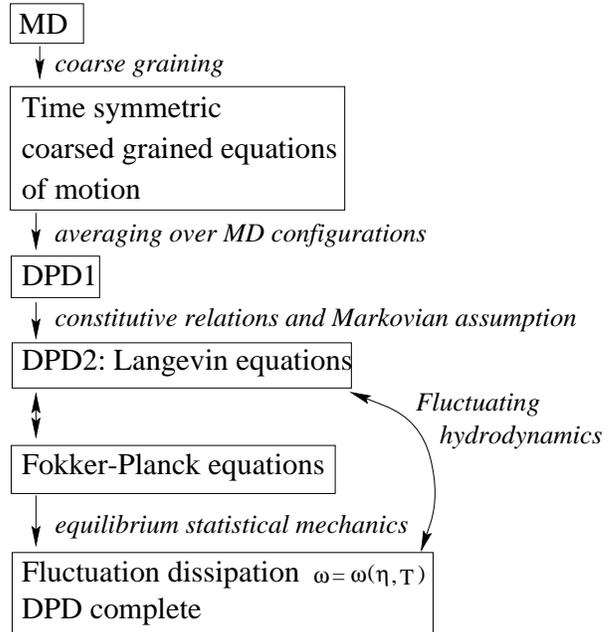,width=8cm}}}
\caption{\label{fig5}
\protect \narrowtext Outline of the derivation of dissipative particle
dynamics from molecular dynamics as presented in the present paper.
The MD viscosity is denoted by $\eta$ and $\omega$ is the amplitude
of the fluctuating force $\tilde{\bf F}$ as defined 
in \protect Eq.~(\ref{fluc_force})}
\end{figure}
Figure \ref{fig5} illustrates the structure of the theoretical development
of DPD equations from MD as presented in this paper.
The initial coarse graining leads to equations of essentially 
the same structure
as the final DPD equations. However, they are still invariant under time-
reversal.
The label DPD1 refers to Eqs.~(\ref{mass_dpd1}), 
(\ref{momentum4})  and   (\ref{energy_dpd1}), whereas
the DPD2 equations have been supplemented with specific constitutive relations 
both for the non-equilibrium fluxes (momentum and heat) and an  equilibrium 
description of the thermodynamics. These equations are 
Eqs.~(\ref{momentum5}) and (\ref{energy_dpd}) along with Eqs.~(\ref{fd}).
The development we have made which is shown in Fig.~\ref{fig5} does
not claim
to derive the irreversible DPD equations from the reversible ones of
molecular dynamics in a rigorous manner, although it 
does illustrate where the transition takes place with the 
introduction of molecular averages. The kinetic equations of this new
DPD satisfy an 
$H$-theorem, guaranteeing an
irreversible approach to the equilibrium state.
Note that in passing to the time-asymmetric description
by the introduction of the averaged description of \eq{constitutive}),
a time asymmetric non-equilibrium ensemble is required~\cite{oppvc94}.

This is the first time that
any of the various existing 
mesoscale methods have been put on a firm `bottom up'
theoretical foundation, a development which brings
with it numerous new insights as well as practical advantages.
One of the main virtues of this procedure 
is the capability it provides to choose one or more coarse-graining 
lengthscales to suit the particular modeling problem at hand.
The relative scale between molecular dynamics and the chosen dissipative
particle dynamics, which may be defined 
as the ratio of their number densities 
$\rho_{\text{DPD}}/\rho_{\text{MD}}$,
is a free parameter within the theory.
Indeed, this rescaling may be viewed as a renormalisation group procedure
under which the fluid viscosity remains constant:
since the conservation laws hold exactly
at every level of coarse graining, the result of doing two rescalings,
say from MD to DPD$\alpha$ and from DPD$\alpha$ to DPD$\beta$, 
is the same as doing just one with a larger ratio, i.e.
$\rho_{\text{DPD}\beta}/\rho_{\text{MD}} = (\rho_{\text{DPD}\beta}/
\rho_{\text{DPD}\alpha})(\rho_{\text{DPD}\alpha}/\rho_{\text{MD}})$. 
% Indeed, the asymptotic limit of infinitely large (or
% indefinitely iterated) coarse-grained rescaling applied to our 
% single-component
% DPD fluid picks out the macroscopic
% Navier-Stokes equations of continuum fluid dynamics as a fixed
% point of this renormalisation group.
% Provided the system at hand  as well as its boundaries 
% exhibit scale invariance this procedure should give rise 
% to corresponding scaling laws.

The present coarse graining scheme is not limited to hydrodynamics.
It could in principle be used to rescale the local description
of any quantity of interest. However, only for 
%NB In view of Giannis findings with Saoro: 'locally' is new
locally
conserved quantities will
the DP particle interactions take the form of surface terms as here,
and so it is unlikely that the scheme will produce a useful
description of non-conserved quantities. 

In this context, we note that the bottom-up approach to fluid
mechanics presented here may throw new light on
aspects of the problem of homogeneous and inhomogeneous
turbulence. Top-down multiscale methods and, to a more limited extent,
ideas taken from renormalisation group theory have been applied 
quite widely in recent years to provide insight into the nature of 
turbulence~\cite{frisch95,bensoussan78}; one might expect an
alternative perspective 
to emerge from a fluid dynamical theory originating at
the microscopic level, in which the central 
relationship between conservative
and dissipative processes is specified in a more fundamental manner.
From a practical point of view it is noted that, 
since the DPD viscosity is the same as the viscosity emergent from the
underlying MD level, it
may be treated as a free parameter in the DPD model, and thus high 
Reynolds numbers may be reached. In the $\eta \rightarrow 0$
limit the model thus represents 
a potential tool for hydrodynamic simulations of turbulence. 
However, we have not investigated the potential numerical 
complications of this limit.

The dissipative particle dynamics which we have derived is formally
similar to the conventional version,
incorporating as it does 
conservative, dissipative and fluctuating forces. The
interactions are pairwise, and conserve mass and momentum as well as
energy.
However, now all these forces have been derived from the underlying 
molecular dynamics. The conservative and dissipative forces arise
directly from the hydrodynamic 
description of the molecular dynamics and 
the properties of the fluctuating forces 
are determined via a fluctuation--dissipation relation.

The simple hydrodynamic description of the molecules chosen 
here is not a necessary requirement.
Other choices for the average of the  general momentum and energy 
flux tensors
Eqs.~(\ref{enrgy_flux}) and (\ref{momentum_flux}) 
may be made and we hope these will be explored in
future work. 
More significant is the fact that our 
analysis permits the introduction of specific
physicochemical interactions at the mesoscopic level, together with
a well-defined scale for this mesoscopic description.

While the Gaussian basis we used for the sampling functions is an
arbitrary albeit convenient choice, the Voronoi geometry itself emerged  
naturally from the requirement that all the MD particles be fully
accounted for. Well defined procedures already 
exist in the literature for the computation of
Voronoi tesselations~\cite{guibas92} and so algorithms 
based on our model are not computationally
difficult to implement. Nevertheless, it should be appreciated that 
the Voronoi construction represents a significant
computational overhead. This overhead is of order
$N\log N$, a factor $\log N$ larger than the 
most efficient multipole 
methods in principle 
available for handling the particle interactions in 
molecular dynamics. 
However, the prefactors are likely to be much larger in the 
particle interaction case.
 
Finally we note the formal similarity of the present
particulate description 
to existing continuum fluid dynamics methods 
incorporating adaptive meshes, 
which start out from a top-down or 
macroscopic description. These top-down approaches include in particular 
smoothed particle hydrodynamics~\cite{monaghan92} and finite-element 
simulations.
In these descriptions too the computational method is based 
on tracing the motion of elements of the fluid on the basis of the 
forces acting between them~\cite{boghosian98}.  
However, while such top-down computational 
strategies depend on a 
macroscopic and purely phenomenological fluid description,
the present approach rests on a {\em molecular} basis.

\acknowledgments 
It is a pleasure to thank Frank Alexander, Bruce Boghosian and Jens
Feder for many helpful and stimulating discussions.  We are grateful
to the Department of Physics at the University of Oslo and
Schlumberger Cambridge Research for financial support which enabled
PVC to make several visits to Norway in the course of 1998; and to
NATO and the Centre for Computational Science at Queen Mary and
Westfield College for funding visits by EGF to London in 1999 and
2000.

% \bibliographystyle{../../bibliography/prsty}
%  \bibliography{../../bibliography/all}

%\appendix
%\section{Mass fluctuations}
%\label{appendix1}
\end{multicols}
%\newpage

\end{document}